\documentclass[traditabstract]{aa}

\usepackage{graphicx}
\usepackage{txfonts}
\usepackage{natbib}
\bibpunct{(}{)}{;}{a}{}{,}

\usepackage{hyperref}
\hypersetup{colorlinks=true,
            citecolor=blue,
            linkcolor=blue}

\def\conj#1{{#1}^\star}
\def\Frac#1#2{{{\displaystyle\strut#1}\over{\displaystyle\strut#2}}}
\def\vv#1{\vec{#1}}
\def\be {\begin{equation}}
\def\ee {\end{equation}}
\def\vR {\vv{R}}
\def\vr {\vv{r}}
\def\ur {\vv{\hat r}}
\def\dvr{\vv{\dot r}}
\def\ddvr{\vv{\ddot r}}
\def\vi {\vv{I}}
\def\vj {\vv{J}}
\def\vk {\vv{K}}
\def\bbi {\vv{i}}
\def\bbj {\vv{j}}
\def\bbk {\vv{k}}

\def\ii {\mathrm{i}}
\def\M {m_0}
\def\m {m}
\def\om {\Omega}
\def\lv {f}
\def\av {v}

\def\tt {\{t\}}
\def\ep {\mathrm{e}}
\def\gam {\gamma}
\def\dgam {2 \gam}
\def\sig {\omega}

\def\eq {{e}}
\def\xix{\zeta}
\def\phid{\delta}

\def\hf{h_\mathrm{f}}
\def\ke{k_\mathrm{e}}
\def\kf{k_\mathrm{f}}
\def\taua{\tau_e}
\def\taub{\tau}
\def\tauv{\tau_v}
\def\tauc{\tau_c}
\def\mue{\mu}
\def\mug{\mu_0}
\def\nue{\eta}

\def\crm{\cr\noalign{\medskip}}

\def\Z{Z}
\def\AA{{\cal A}}
\def\BB{{\cal B}}
\def\CC{{\cal C}}
\def\KK{{\cal K}}
\def\Ze{Z^{\eq}}
\def\ttheta{\tilde\theta}
\def\torque{T}
\def\llambda{\nu}

\def \llabel#1{\label{#1}}

\begin{document}

\title{Deformation and tidal evolution of close-in planets \\ and satellites 
using a Maxwell viscoelastic rheology}
\titlerunning{Deformation and tidal evolution using a Maxwell rheology}


\author{
Alexandre C.M. Correia\inst{1,2}
\and Gwena\"el Bou\'e\inst{2}
\and Jacques Laskar\inst{2}
\and Adri\'an Rodr\'iguez\inst{3}
}

 
\institute{Departamento de F\'isica, I3N, Universidade de Aveiro, Campus de
Santiago, 3810-193 Aveiro, Portugal
  \and 
ASD, IMCCE-CNRS UMR8028,
Observatoire de Paris, UPMC,
77 Av. Denfert-Rochereau, 75014 Paris, France  
\and 
Instituto de Geoci\^encias e Ci\^encias Exatas, UNESP, Av. 24-A 1515, CEP 13506-900, Rio Claro, SP, Brazil.
}

\date{Received ; accepted To be inserted later}

\abstract{
In this paper we present a new approach to tidal theory. 
Assuming a Maxwell viscoelastic rheology, we compute the instantaneous deformation of celestial bodies using a differential equation for the gravity field coefficients.
This method allows large eccentricities and it is not limited to quasi-periodic  perturbations.  
It can take into account an extended class of perturbations, including chaotic motions and transient events.
We apply our model to some already detected eccentric hot Jupiters and super-Earths in planar configurations.
We show that when the relaxation time of the deformation is larger than the orbital period, spin-orbit equilibria arise naturally at half-integers of the mean motion, even for gaseous planets.
In the case of super-Earths, these equilibria can be maintained for very low values of eccentricity.
Our method can also be used to study planets with complex internal structures and other rheologies.
}


   \maketitle
%



\section{Introduction}

The occurrence, on most open ocean coasts, of high sea tide at about the time of the Moon's passage across the meridian, gave ancient observers the idea that our satellite exerts an attraction on the water, although the occurrence of a second high tide when the Moon is on the opposite meridian was a great puzzle.
The correct explanation for tides was first indicated by \citet{Newton_1760} in the {\it Philosophi{\ae} Naturalis Principia Mathematica}.
They result from a differential effect of the gravitational force acting in accordance with laws of mechanics, since this force is not constant across the diameter of the Earth.
\citet{Kant_1754} suggested that the Moon not only pulls the Earth, but also exerts a retarding torque upon its surface, this torque slowing down the Earth's rotation until terrestrial days become as long as lunar months.

The estimates for the tidal deformation of a body are based on a very general formulation of the tidal potential, initiated by \citet{Darwin_1879a}.
Assuming a homogeneous Earth consisting of an incompressible fluid with constant viscosity, Darwin derived a tide-generated disturbing potential expanded into a Fourier series.
Thus, he was able to obtain the tidal contribution to the spin and orbital evolution, confirming Kant's prediction \citep{Darwin_1880}.
Ever since, many authors improved Darwin's work, leading to a better understanding of this problem \citep[e.g.,][]{Kaula_1964, MacDonald_1964, Alexander_1973, Singer_1968, Mignard_1979, Efroimsky_Williams_2009, Ferraz-Mello_2013}.

All previous studies have shown the existence of a stationary solution for the rotation rate, which is synchronous when the two bodies move in circular orbits, but becomes super-synchronous for elliptical orbits.
Standard tidal models consider an elastic tide and delay the tidal bulge by an ad hoc
phase lag in order to take into account the body anelasticity \citep[e.g.,][]{Kaula_1964,MacDonald_1964}.
However, these models are unable to explain the dependency of phase lag with the tidal frequency, which is a crucial point when we aim to study the long-term secular evolution of the planets.

If one adopts a linear model where the phase lag is proportional to the tidal frequency, i.e, a {\it viscous}Ê model, the ratio between the orbital and rotational periods have an excess proportional to the square of the eccentricity, an equilibrium often called {\it pseudo-synchronous} rotation \citep[e.g.,][]{Mignard_1979}.
However, this prediction is not confirmed by observations of the rotation of most planetary satellites (which are synchronous in non-eccentric orbits) or of the planet Mercury.
One way to obtain a synchronous stationary solution is to assume that an additional torque, due to a permanent equatorial asymmetry (a quadrupolar approximation for the body's figure), is acting on the body to counterbalance the tidal torque \citep{Colombo_1965, Goldreich_Peale_1966}.
The stationary solutions obtained from the combination of both torques are called spin-orbit resonances, for which the synchronous motion is a particular case. 

A more realistic approach to deal with the dependency of the phase lag with the tidal frequency is to assume a {\it viscoelastic} rheological model.
One of the simplest models of this kind is to consider that the planet behaves like a Maxwell material, i.e., the material is represented by a purely viscous damper and a purely elastic spring connected in series \citep[e.g.,][]{Turcotte_Schubert_2002}.
In this case, the planet can respond as an elastic solid or as a viscous fluid, depending on the frequency of the perturbation.
Viscoelastic rheologies have been used recently \citep[e.g.,][]{Henning_etal_2009, Remus_etal_2012b}, since they are able to reproduce the main features of tidal dissipation, including the {\it pseudo-synchronous} rotation or the spin-orbit equilibria \citep[e.g.,][]{Storch_Lai_2014}
\citep[For a review of the main viscoelastic models see][]{Henning_etal_2009}.

\citet{Efroimsky_2012} describes 
the rheology of a planet using an empirical power scaling law for the phase lag which is consistent with the accumulated geophysical, seismological, and geodetic observational data.
It is shown that
the best fitted laws are in conformity with Andrade's model \citep{Andrade_1910}, where the phase lag is inversely proportional to the tidal frequency.
The Andrade model can be thought of as the Maxwell model equipped with an extra
term describing hereditary reaction of strain to stress \citep{Efroimsky_2012}.
However, Andrade's model is extremely complex and the main features of tidal evolution are similar to other viscoelastic models.

\citet{Ferraz-Mello_2013} abandons the potential description introduced by Darwin, and proposes to study tides directly from the deformation law of the planet.
For that purpose, he adopts a simple Newtonian creep model for the surface deformation, where  the distance to the equilibrium is considered as proportional to the stress.
As a result, the planet's surface tends to the equilibrium spheroid, but not instantaneously.
The resulting stationary rotations have an average excess velocity which depends on a critical frequency, the relaxation factor, which is inversely proportional to the viscosity of the body.
Moreover, the dissipation in the {\it pseudo-synchronous} solution presents two regimes of variation, depending on the tidal frequency:
In the inviscid limit, it is roughly proportional to the frequency (as in standard theories), but that behavior is inverted when the viscosity is high and the tidal frequency larger than the critical frequency, as in the model proposed by \citet{Efroimsky_2012}.

An important setback in \citet{Ferraz-Mello_2013} model is the absence of an elastic tide. 
Indeed, when the tidal frequency is larger than the critical frequency, the phase lag tends to $90^\circ$, which is in contradiction with the observations done for most rocky planets.
In order to conciliate the theory and the observed tidal bulges, we have to assume that the tide is not restricted to the component due to the creeping of the body
under the tidal action, but has also a pure elastic component. 
This supposition was already present in Darwin's original work \citep{Darwin_1879a}, who realized that {\it viscoelastic} deformations of the planet's surface could be derived from the Newtonian creep law.

In this paper we pursue in the same direction of \citet{Ferraz-Mello_2013}, but we adopt a deformation law for the planet that is compatible with the viscoeslatic rheology of a Maxwell body (section~\ref{TheModel}).
We then compute the general solution for the deformation of the planet (section~\ref{gensol}), and use it to obtain the tidal evolution of the planet (section~\ref{tidalevol}).
We also identify the equilibrium positions for the tidal torque (section~\ref{sores}), and inspect the effect from some libration of the rotation rate around these equilibria (section~\ref{libsec}).
We finally perform numerical simulations for some known close-in exoplanets (section~\ref{appexo}) and discuss the results in the last section.


\section{Model}

\llabel{TheModel}

\begin{figure}
\centering
\includegraphics[width=8cm]{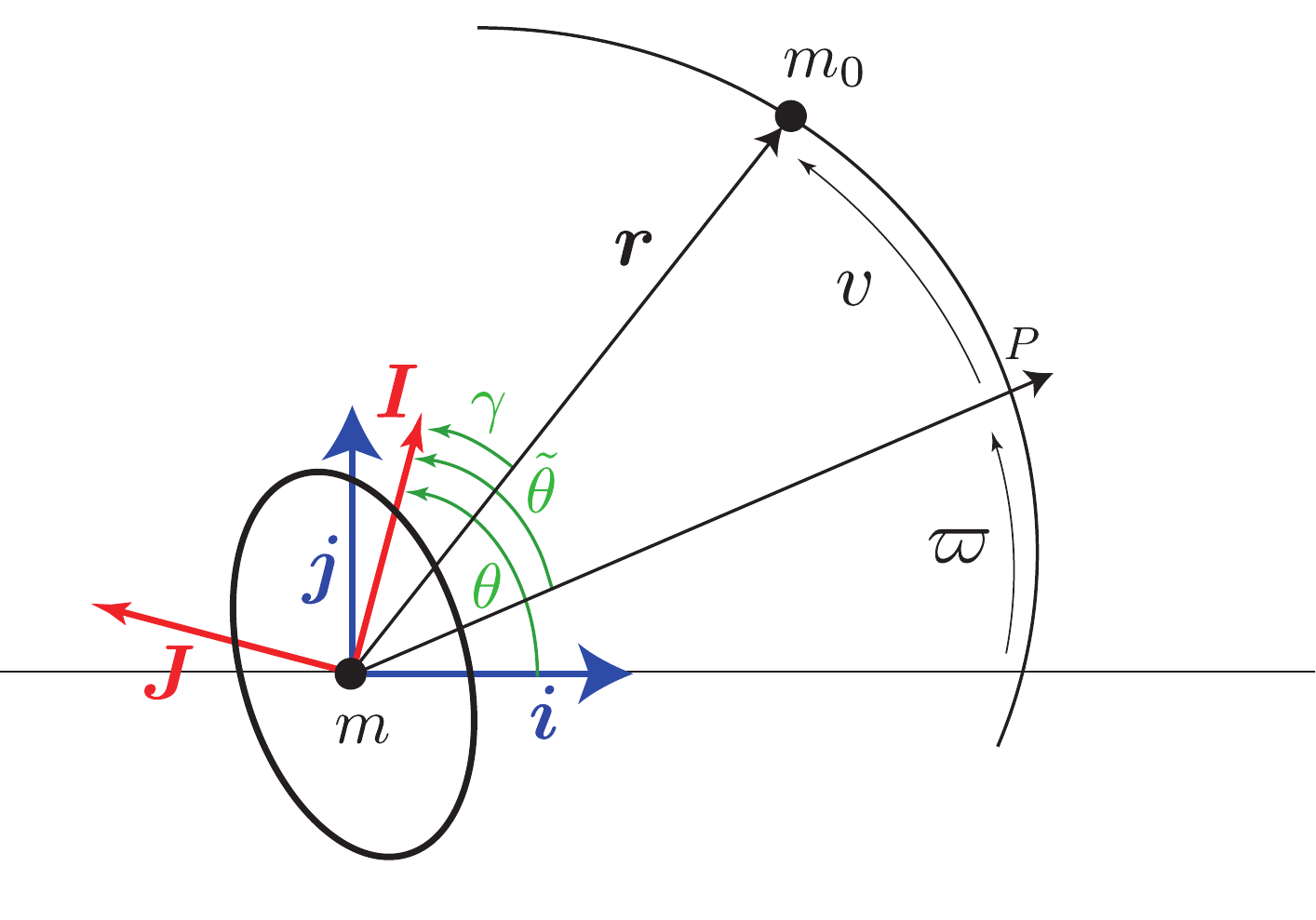}
\caption{Reference angles and notations. ($\bbi,\bbj,\bbk$)  is a fixed inertial reference frame, while ($\vi,\vj,\vk$) is a reference frame fixed in the planet. $\bbk = \vk$ are orthogonal to the orbit, and thus not shown. $\vr$ is the radius vector from $\m$ to $\M$, $P$ is the direction of  pericenter, $\varpi$ is the longitude of pericenter, and $\av$ is the true anomaly. The rotation angle $\theta$ is referred to the fixed reference 
direction $\bbi$. We also use the angles $ \ttheta = \theta -\varpi$, $\lv=\av+\varpi$, and
$\gamma = \theta-f$.  \llabel{Fignot} }
\end{figure}

We consider a system consisting of a central star of mass $\M$, and a companion planet of mass $\m$.
The planet is considered an oblate ellipsoid with gravity field coefficients given by $J_2$, $C_{22}$ and $S_{22}$, sustained by the reference frame ($\vi,\vj,\vk$), 
where $\vk$ is the axis of maximal inertia (Fig.\,\ref{Fignot}).
We furthermore assume that the spin axis of the planet, with rotation rate $\om$, is also along $\vk$ (gyroscopic approximation), and that $\vk$ is orthogonal to the orbital plane (which corresponds to zero obliquity).
The gravitational potential of the planet at the star is then given by \citep[e.g.,][]{Correia_Rodriguez_2013}

\begin{eqnarray}
V (\vr) \!&=&\! - \frac{G \m}{r}
- \frac{G \m R^2 J_2}{2 r^3} 
- \frac{3 G \m R^2}{r^3}  \big(  C_{22} \cos \dgam  - S_{22} \sin \dgam \big) \ , \llabel{130528a} 
\end{eqnarray}
where
\be
\cos \dgam =  (\vi \cdot \ur)^2 - (\vj \cdot \ur)^2
\quad \mathrm{and} \quad 
\sin \dgam = - 2 (\vi \cdot \ur) (\vj \cdot \ur) \ ,
\ee
$G$ is the gravitational constant, $R$ is the mean planet radius,  $\vr$ is the position of the star with respect to center of mass of the planet, and $\ur = \vr / r $ is the unit vector.
We neglect terms in $(R/r)^3$ (quadrupolar approximation).

\subsection{Deformation of the planet}

The mass distribution inside the planet, characterized by the gravity field coefficients, is a result of the forces acting on it, that is, self gravity and the planet's response to any perturbing potential, $V_p $.
Here, we consider that the planet deforms under the action of the centrifugal and tidal potentials.
Thus, on the planet's surface, $\vR$, 
the non-spherical contribution of the perturbing potential is given by \citep[e.g.,][]{Correia_Rodriguez_2013}
\be
V_p (\vR) = \frac{1}{2} \om^2 (\vk \cdot \vR)^2 - \frac{3 G \M}{2 r^3} (\ur \cdot \vR)^2  \ . \llabel{130529a}
\ee

We let $\xix (\vR)$ be the radial deformation of the free surface.
In the case of a homogeneous incompressible {\it elastic} solid sphere, the equilibrium surface deformation $\xix_\eq(\vR)$ is given by \citep{Kelvin_1863b,Kelvin_1863a,Love_1911}
\be
\xix_\eq (\vR) = - h_2 V_p (\vR) / g \ , \llabel{131021a}
\ee
where $ g = G \m / R^2$ is the surface gravity, and
\be
h_2 = \frac{5}{2} \left(1+\frac{19 \mue}{2 g \rho R}\right)^{-1}  \llabel{131021b}
\ee
is the second Love number associated with the radial displacement, $\rho$ the density, and $\mue$ the rigidity (or shear modulus). \citet{Darwin_1879a} extended the results to a homogeneous incompressible {\it viscous} body and obtained the differential equation
\be
\xix + \tauv \dot \xix = - \hf V_p / g \ ,  \llabel{131021c}
\ee
where $\hf=5/2$ is the fluid Love number\footnote{
The fluid Love numbers depend on the internal differentiation of the bodies. For a homogeneous incompressible viscous sphere we have $\hf = 5/2$ and $\kf = \hf - 1 = 3/2$, 
but more generally the fluid Love numbers can be obtained from the Darwin-Radau equation \citep[e.g.,][]{Jeffreys_1976}: 
$ \hf = 5 (1+[5/2-15C/(4\m R^2)]^2)^{-1} $.},
\be
\tauv = \frac{19 \nue}{2 g \rho R} = \frac{38 \pi R^2 \nue}{3 g \m} \llabel{130524b}
\ee
is the fluid relaxation time, and $\nue$ is the viscosity.
\citet{Darwin_1879a} also realized that {\it viscoelastic} deformations can easily be derived
from expression (\ref{131021c}) using the substitution
\be
\frac{1}{\nue} \rightarrow \frac{1}{\mue} \left(\frac{1}{\taua}+\frac{d}{dt}\right) \ ,  \llabel{131021d}
\ee
where $\taua = \nue / \mue$ is the elastic or Maxwell relaxation time.
As a result, we obtain \citep{Darwin_1879a}
\be
\xix + \taub \dot \xix = - \hf (V_p + \taua \dot V_p) / g \ , \quad \mathrm{with} \quad \taub = \tauv+\taua \ .  \llabel{131021e}
\ee

Finally, the deformation $\xix$ of the body generates an additional potential $V'(\vr)$ given by
\be
V'(\vr) = - \int \frac{G \rho \xix(\vR')}{|\vr-\vR'|} \, d \vR'
 \ .  \llabel{131021f}
\ee
In the case of quadrupolar deformations, as considered here (Eq.\,\ref{130528a}), the additional potential exterior to the body ($r \ge R$) is
\be
V'(\vr) = - \frac{3}{5} g \left(\frac{R}{r}\right)^3 \xix(\vR)
 \ .  \llabel{131021g}
\ee
Thus, at the mean radius ($r=R$), the additional potential must satisfy
\be
V' + \taub \dot V' = \kf (V_p + \taua \dot V_p)  \ ,  \llabel{130524z}
\ee
where $\kf = 3/2$ is the fluid second Love number for potential\footnotemark[\value{footnote}].
The equilibrium potential $V_e$ is obtained for static perturbations or instantaneous response ($\taub=\taua=0$)
\be
V_e = \kf V_p = \frac{\kf}{2} \om^2 (\vk \cdot \vR)^2 - \kf \frac{3 G \M}{2 r^3} (\ur \cdot \vR)^2  \ . \llabel{130528z}
\ee
The instantaneous variation of the additional potential $V'$ that results from the perturbation $V_p$, can then be obtained from the equilibrium potential $V_e$ through expression (\ref{130524z}) as
\be
V' + \taub \dot V' =  V_e + \taua \dot V_e  \ .  \llabel{130524f}
\ee
We note that this {\it viscoelastic} deformation law is also fully compatible with a {\it viscous} rheology (Eq.\,\ref{131021c}), provided that we set $\taua=0$.
Previous expression could also have been obtained in the Fourier domain (see appendix~\ref{apenA}).

\subsection{Equations of motion}

The force between the star and the planet is easily obtained from the gravitational potential  of the system (Eq.\,\ref{130528a}) as $\vv{F} = - \M \times \nabla V (\vr) $.
From this force we can directly obtain the orbital evolution of the system $\ddvr =  \vv{F} / \beta $, and the spin evolution $ \ddot\theta = - (\vr \times \vv{F}) \cdot \vk / C $ (using the conservation of the total angular momentum of the system), where $\beta = \M \m/(\M + \m)$ is the reduced mass of the system, and $ C $ is the principal inertia moment of the planet.
Expressing $ \ur = (\cos \lv, \sin \lv) $ and $ \vi = (\cos \theta, \sin \theta) $, where $\lv$ is the true longitude and $\theta$ is the rotation angle, we thus have
\begin{eqnarray}
\ddvr = & - & \frac{\mug}{r^2} \ur 
- \frac{3 \mug R^2}{2 r^4} J_2 \, \ur  
- 
\frac{9 \mug R^2}{r^4} \left[ 
C_{22}  \cos \dgam - S_{22}  \sin \dgam
 \!  \phantom{\frac{}{}} \right] \ur   \crm 
 & + & \frac{6 \mug R^2}{r^4}  \left[ C_{22} \sin \dgam + S_{22} \cos \dgam
  \! \phantom{\frac{}{}} \right] \vk \times \ur
\ , \llabel{130104b}
\end{eqnarray}
and
\begin{eqnarray}
\ddot \theta \!&=&\! - \frac{6 G \M \m R^2}{C r^3}  \left[ 
C_{22} \sin \dgam + S_{22} \cos \dgam
\! \phantom{\frac{}{}} \right]  \ , \llabel{130104d}
\end{eqnarray}
where $\mug = G (\M + \m) $, and $\gam = \theta - \lv$ (Fig.\,\ref{Fignot}).

Because the planet is not rigid and can be deformed under the action of a perturbing potential, the gravity field coefficients $J_2$, $C_{22}$ and $S_{22}$ are not constant.
Indeed, according to the deformation law given by expression (\ref{130524f}) we have
\be
J_2 + \taub \dot J_2 = J_2^\eq + \taua \dot J_2^\eq \ , \llabel{130107p} 
\ee
\be
C_{22} + \taub \dot C_{22} = C_{22}^\eq + \taua \dot C_{22}^\eq \ , \llabel{130107c} 
\ee
\be
S_{22} + \taub \dot S_{22} = S_{22}^\eq + \taua \dot S_{22}^\eq \ . \llabel{130107q} 
\ee

The equilibrium values for each coefficient are easily obtained comparing expressions (\ref{130528a}) and (\ref{130528z}), which gives \citep{Correia_Rodriguez_2013}
\be
J_2^\eq = \kf \left[ \frac{\om^2 R^3}{3 G \m} + \frac{1}{2} \frac{\M}{\m} \left(\frac{R}{r} \right)^3  \right]  \llabel{130104e} \ ,
\ee
\be
C_{22}^\eq =  \frac{\kf}{4} \frac{\M}{\m} \left(\frac{R}{r}\right)^3 
\cos \dgam
\ , \llabel{130104f} 
\ee
\be
S_{22}^\eq = - \frac{\kf}{4} \frac{\M}{\m} \left(\frac{R}{r}\right)^3  
\sin \dgam
\ , \llabel{130104g} 
\ee
and for their time derivatives, respectively
\be
\dot J_2^{\eq} = \kf \left[ \frac{2 \om R^3}{3 G \m} \dot\om - \frac{3}{2} \frac{\M}{\m} \left(\frac{R}{r}\right)^3 \frac{\dvr \cdot \vr}{r^2}  \right]  \llabel{130107e} \ ,
\ee
\be
\dot C_{22}^{\eq} = - \frac{\kf}{4} \frac{\M}{\m} \left(\frac{R}{r}\right)^3 \left[ 3 \frac{\dot r}{r}  \cos \dgam + 2 \dot \gam \sin \dgam  \right]  \ , \llabel{130107f} 
\ee
\be
\dot S_{22}^{\eq} = \frac{\kf}{4} \frac{\M}{\m} \left(\frac{R}{r}\right)^3 \left[ 3 \frac{\dot r}{r}  \sin  \dgam - 2 \dot \gam \cos \dgam  \right]  \ . \llabel{130107g} 
\ee
It should be noted that for the equilibrium solution (Eqs.\,\ref{130104f}, \ref{130104g}), 
we have $\ddot \theta = 0 $ (Eq.\ref{130104d}).

\section{General solution}
\llabel{gensol}

Introducing the complex notations
\be
\Z = C_{22} - \ii \,S_{22}\quad\mathrm{and} \quad \Ze= C^{\eq}_{22} - \ii \,S^{\eq}_{22} \ ,
\ee
we can rewrite equations  (\ref{130107c}) and (\ref{130107q}) together as 
\be
\Z + \taub \dot\Z = \Ze + \taua \dot\Ze \ .
\llabel{eqzez}
\ee
The solution of equation (\ref{eqzez}) without the second hand member is 
$ \Z = \Z_0 \ep^{-t/\taub} \ , $
and the variation of the constant $\Z_0$ gives the general solution of equation (\ref{eqzez})
\be
\Z (t) = Ê\frac{\taua}{\taub} \, \Ze (t) + \frac{1}{\taub} \Big(1-\frac{\taua}{\taub} \Big) \int_0^t \Ze (t') \, \ep^{(t'-t)/\taub} d t' +  \CC \, \ep^{-t/\taub}  ,
\llabel{140506a}
\ee
where $\CC$ is an integration constant. 
Although one can search for a solution of this equation in the precise setting of this paper, 
it is much more efficient  to solve it first in a very general setting where $\Ze$ is a general 
almost periodic function, that is an expression of the form 
\be
 \Ze (t) = \sum_{k}  \beta_{k} \, \ep^{\ii  \sig_k t}  \ , 
 \llabel{eqze}
 \ee
where $\beta_k$ are constants, and $\sig_k$ are a numerable set of fixed frequencies. The result is
\be
\Z(t)=\sum_k \beta_{k} \frac{ 1 + \ii \taua \sig_k}{ 1 + \ii \taub \sig_k} \, \ep^{\ii  \sig_k t}  + \CC \, \ep^{-t/\taub} \ .
 \llabel{eqzz}
\ee
The first term is the stationnary mode and the second one 
the transient mode that will decay rapidly to zero with time. In
general, the constant $\CC$ in equation (\ref{eqzz}) is different from
that of equation (\ref{140506a}).
It should be noted that here we do not have small divisors problems as the denominators $1 + \ii \taub\sig_k$
never vanish. A constant term $\beta_{k_0}$ in $\Ze$
(corresponding to $\sig_{k_0}=0$) will add an equivalent constant contribution $\beta_{k_0}$ to $\Z$.

\section{Tidal evolution}
\llabel{tidalevol}
From the general solution of equation (\ref{eqzez}), we can study more specific solutions. 
For this, we will look to the explicit expression of $\Ze$. Using expressions
(\ref{130104f}) and (\ref{130104g}), we can write 
\be
\Ze = \AA \left(\frac{a}{r}\right)^3 \ep^{\ii 2 (\ttheta-\av)} \ , \quad \hbox{with} \quad \AA=  \frac{\kf}{4} \frac{\M}{\m} \Big(\frac{R}{a}\Big)^3 \ , 
\ee
where $a$ is the semi-major axis, $\ttheta=\theta-\varpi$, $\varpi$ is the longitude of the pericenter, and $\av$ is the true anomaly (Fig.\,\ref{Fignot}). 
Using the  expansion of $(a/r)^3\exp(-2\ii \av)$ in terms of   Hansen coefficients $X_k^{l,m} (e) $  \citep[e.g.,][]{Hughes_1981} (see also appendix~\ref{appendix.Hansen}), 
\be 
\left( \frac{r}{a} \right)^l \ep^{\ii m \av} =  \sum^{+\infty}_{k=-\infty} X_k^{l,m} (e) \, \ep^{\ii k M} \ , \llabel{061120gb}
\ee
we obtain
 \be
 \Ze = \sum_{k=-\infty}^{+\infty}  \beta_k \, \ep^{\ii (2\ttheta-k M)} \ , \quad \mathrm{with} \quad \beta_k = \AA \, X_k^{-3,2}(e) \ , 
 \llabel{eqzeb}
 \ee
where $M$ is the mean anomaly, and $e$ is the eccentricity.
$\beta_k$ depends only on the orbital parameters that are assumed to be constant over one revolution. 
The evolution of $\theta$ is given by equation (\ref{130104d}), 
but since tidal effects are usually very weak, in a zero order solution, we can also neglect $\ddot \theta$, and assume  that the rotation rate $\om = d \ttheta / dt $ is constant over one orbital revolution.
Thus, we can write

\be
\sig_k = 2 \om - k n = \frac{d}{dt}(2 \ttheta - k M ) \ , \llabel{131105b}
\ee
where $ n = \dot M $, and $\sig_k$ is a constant frequency. As the $\beta_k$ are constant in equation (\ref{eqze}),
we are in a special case of the application of the results of  section \ref{gensol}, and we obtain, after neglecting the 
transtient part of expression (\ref{eqzz}) that decays to zero for $t \gg \taub$,
\be
\Z = \sum_{k=-\infty}^{+\infty}  \beta_k \frac{1 + \ii \taua \sig_k}{1 + \ii \taub \sig_k} \, \ep^{\ii (2\ttheta-k M)}  \ .
\llabel{eqzzb}
\ee


\begin{figure*}
\centering
\includegraphics*[width=\textwidth]{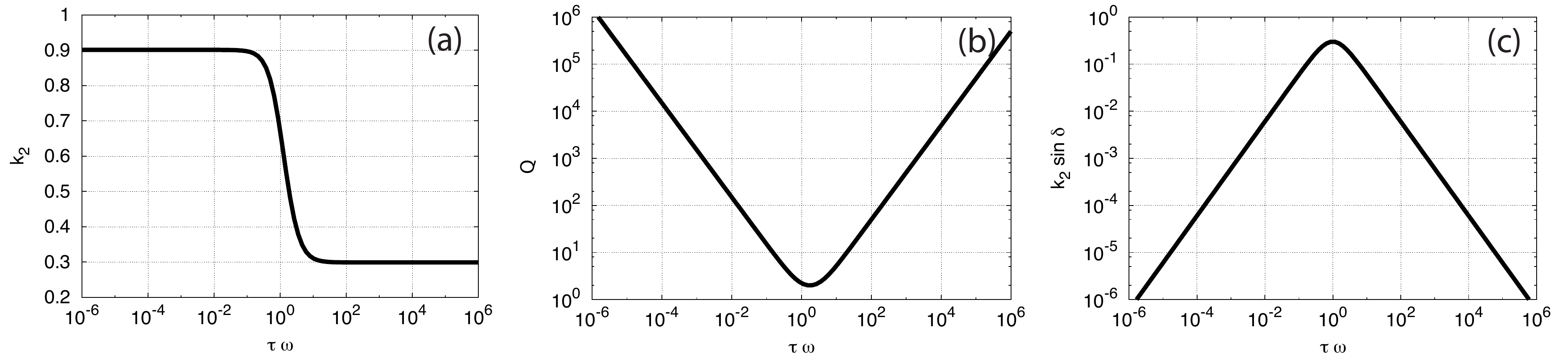}
\caption{Evolution of the second Love number $k_2$, the dissipation $Q$-factor, and phase lag $k_2 \sin \phid$ as a function of the tidal frequency $\sig$ (normalized by the relaxation time $\taub$).
For all parameters there is a transition of regime around $\taub \sig \sim 1$.  
We use here the Earth values for the Love numbers, $\ke = 0.3$ and $\kf = 0.9$ \citep{Yoder_1995cnt}.
\llabel{Fig1} }
\end{figure*}

\subsection{Tidal torque}

We have already seen that for $\Z = \Ze$, $\ddot\theta=0$, as there are no coupling torques in the problem due to alignment of the planet deformation with the star.
It is then useful to use the difference 
\be
\Delta\Z = \Z-\Ze = -\left(1-\frac{\taua}{\taub} \right) \ep^{\ii 2 \ttheta} 
\sum_{k=-\infty}^{+\infty}  \beta_k \frac{\ii \taub \sig_k}{1 + \ii \taub \sig_k} \, \ep^{-\ii k M}  \ .
\llabel{deltaz}
\ee
Denoting 
\be
\BB = \frac{6 G \M \m R^2}{C a^3} \ ,
\ee
we obtain the variations of $\ddot\theta$ in the first order solution
\be
\ddot \theta = \BB \left(\frac{a^3}{r^3}\right) \Im \left( \Delta\Z \, \ep^{-\ii 2(\ttheta - \av)}\right)
= \frac{\BB}{\AA} \Im \left( \Delta\Z \, \conj{\Ze}\right)\ .
\llabel{ddottheta}
\ee
where $\Im (z) $ denotes the imaginary part of $z$ and $\conj{z}$ its complex conjugate. We thus see from expression (\ref{deltaz}) that $\theta$ disappears from 
the expression of $\ddot \theta$.
After an expansion in Hansen coefficients  of expression (\ref{ddottheta}), we obtain
an explicit expression of $\ddot \theta$ with respect to the mean anomaly $M$
\begin{eqnarray}
\ddot \theta &=& - \KK \, \Im\left( \left(\sum_{k=-\infty}^{+\infty}  X_k^{-3,2}  \frac{\ii \taub \sig_k}{1 + \ii \taub \sig_k}\ep^{-\ii k M}\right)\left(\sum_{l=-\infty}^{+\infty}X_l^{-3,2}\ep^{\ii l M}\right)\right) \crm 
&=& 
- \KK  \sum_{j=-\infty}^{+\infty} \sum_{k=-\infty}^{+\infty} X_k^{-3,2} X_{j+k}^{-3,2} \, \Im\left(\frac{\ii \taub \sig_k}{1 + \ii \taub \sig_k}\ep^{\ii j M} \right) \ ,
\llabel{ddotthetac}
\end{eqnarray}
where 
\be
\KK = \AA \BB \left(1 - \frac{\taua}{\taub} \right) = \kf \frac{3 G \M^2 R^5}{2 C a^6} \left( 1 - \frac{\taua}{\taub} \right) \ .
\ee
The first order secular evolution of $\ddot\theta$ over one orbital period is easily obtained by averaging expression (\ref{ddotthetac}) over $M$, which is equivalent to keep only the terms with $j=0$.
We thus get $\langle \ddot \theta \rangle_M = \torque$, with the secular torque $\torque$ given by
\be
\torque =  - \KK \sum_{k=-\infty}^{+\infty} \left(X_k^{-3,2}\right)^2 \frac{\taub\sig_k}{1 + \taub^2\sig_k^2} \ . \llabel{ddotthetaM}
\ee


\subsection{Relation with classical notations}

Expressions (\ref{eqzzb}) and (\ref{ddotthetaM}) can be related to classical tidal parameters, by transforming the complex notation in a real amplitude and a phase lag
\begin{eqnarray}
\Z = \frac{1}{4} \frac{\M}{\m} \Big(\frac{R}{a}\Big)^3 \sum^{+\infty}_{k=-\infty} X_k^{-3,2} (e) \, k_2(\sig_k) \, \ep^{\ii (2\ttheta-k M -\phid_k)  } \ , \llabel{130530a} 
\end{eqnarray}
and
\be
\torque =  - \frac{3 G \M^2 R^5}{2 C a^6} \sum_{k=-\infty}^{+\infty}  \left(X_k^{-3,2}\right)^2 k_2(\sig_k) \sin \phid_k \ , \llabel{140506b}
\ee
with 
\be \tan \phid_k = \frac{(\taub-\taua) \sig_k}{1+ \taub \taua \sig_k^2} \ , \llabel{131004z} \ee
and
\be
k_2 (\sig_k) = \kf \sqrt{\frac{1+ (\taua \sig_k)^2}{1+ (\taub \sig_k)^2}} \ .
\llabel{130923c}
\ee

We then conclude that the $C_{22}$ and $S_{22}$ are shifted by an angle $\phid_k$ with respect to the perturbation, 
and weakened by a factor of $k_2(\sig_k)/\kf$ (Fig.\,\ref{Fig1}).
The phase lag $\phid_k$ (Eq.\,\ref{131004z}) is zero for $\sig_k = 0$ and $\sig_k \rightarrow \infty$, and its maximal value is obtained at the frequency $ \sig_k = (\taub \taua)^{-1/2} $.
The  amplitude $ k_2 (\sig_k) $ is monotonically decreasing with $\sig_k$  with 
\be
 \kf =k_2 (0) \ge   k_2 (\sig_k)  \ge k_2 (\infty)  = \kf \frac{\taua}{\taub} \equiv \ke  \ .\llabel{131007a}
\ee
 We call $\ke$ the {\it elastic} Love number. 
One can consider that the transition between the {\it elastic} and the {\it viscous} regimes 
occurs for $k_2(\sig_0) = (\kf+\ke)/2$. Because of the expression of $ k_2$ (Eq.\,\ref{130923c}), 
we use $k_2(\sig_0) = \sqrt{(\kf^2+\ke^2)/2}$, which gives the very simple transition condition
\be
 \taub \sig_0 = 1 \ .
\ee
If one is able to measure $\ke$, $\kf$ and $\phid_k$ for a planet, it becomes possible to determine the relaxation times $\taub$ and $\taua$.

\subsection{Energy dissipation and orbital evolution}

Consider that $\dot \theta = \om + \delta\om$, where $\om $ is constant and $\delta\om$ a small perturbation of $\om$. 
Neglecting terms of second order, 
the rotational energy dissipated in the planet is then easily obtained as 
\be
\langle \dot E_\mathrm{rot} \rangle_M = \left\langle \frac{d}{dt} \left( \frac{1}{2} C \, \dot\theta^2 \right) \right\rangle_M \approx C \, \om \, \langle \ddot\theta \rangle_M = C \, \om \, \torque \ .
\llabel{131004a}
\ee

The  variation of orbital energy is obtained as the variation of the potential energy (Eq.\,\ref{130528a})
under the variation of the potential coefficients, that is 
\be
\dot E_\mathrm{orb} = \M\frac{\partial V}{\partial t}  = - \frac{G \M \m R^2}{2a^3}\left[\frac{a^3}{r^3} \dot J_2  + 6\Frac{a^3}{r^3} \, \Re\left( \dot\Z \ep^{i2(\lv-\ttheta)}\right)\right] \ ,
\ee
where $\Re(z)$ is the real part of the complex number $z$. 
Using the expression of $J_2^{\eq} $ (Eq.\,\ref{130104e}) and section~\ref{gensol},  we have 
\be
\dot  J_2 =  2 \ii \AA  \sum_{k=-\infty}^{+\infty} X_k^{-3,0}kn \Frac{1+ \ii \taua k n }{1+ \ii \taub k n } \ep^{ikM} \ ,
\ee 
and from expression (\ref{eqzzb})
\be
\Frac{a^3}{r^3}\dot\Z \ep^{i2(\lv-\ttheta)} = \ii \AA \left(\sum_{k=-\infty}^{+\infty} X_k^{-3,2} \sig_k\Frac{1+ \ii \taua \sig_k }{1+ \ii \taub \sig_k } \ep^{-ikM}\right) \left(\Frac{a^3}{r^3} \ep^{i2v}\right) \ .
\ee
Finally, after averaging over the mean anomaly $M$, 
\be
\langle \dot E_\mathrm{orb} \rangle_M =  \frac{C \KK}{6}  \sum_{k=-\infty}^{+\infty} \left[ \left(X_k^{-3,0}\right)^2 \! \frac{\taub k^2 n^2 }{1+  \taub^2 k^2 n^2 } + \left(X_k^{-3,2}\right)^2  \!\frac{3 \taub \sig_k^2}{1+  \taub^2 \sig_k^2 } \right] \ .
\ee
The last term results form the contribution of $C_{22}$ and $S_{22}$, while the first term results from the contribution of the $J_2$, so it does not depend on the rotation angle (Eq.\,\ref{130107p}).
We can also easily compute the semi-major axis and eccentricity evolution from the tidal energy \citep[e.g.,][]{Correia_etal_2011} 
\be
\dot a = \frac{2 \dot E_\mathrm{orb}}{\beta a n^2} \ ,  \quad \mathrm{and} \quad 
\dot e = \frac{(1-e^2)}{\beta n a^2 e} \left( \frac{\dot E_\mathrm{orb}}{n} + \frac{\dot E_\mathrm{rot}}{\om \sqrt{1-e^2}} \right) \ . \llabel{140130d} 
\ee

The total energy released inside the body due to tides is then
\be
\dot E = - (\dot E_\mathrm{rot} + \dot E_\mathrm{orb}) \ . \llabel{140130b}
\ee
A commonly used dimensionless measure of the tidal dissipation is the quality factor, 
\be
Q^{-1} = \frac{1}{2 \pi E_0} \oint \dot E \, d t \ , \llabel{131004b}
\ee
where the line integral over $\dot E$ is the energy dissipated during one period of tidal stress, and $E_0$ is the peak energy stored in the system during the same period.
In general, $Q$ is a function of the frequency $\sig_k$, and it can be related to the phase lag through
$Q_k^{-1} = \sin \phid_k$
\citep[][Eq.\,141]{Efroimsky_2012}. In that case, the ratio $k_2/Q_k$ is
directly connected with the tidal torque (Eq.\,\ref{140506b}) and
therefore it is easily measurable
\be
k_2(\sig_k)/Q_k =  k_2(\sig_k) \sin \phid_k = \kf \frac{(\taub-\taua)\sig_k}{1+(\taub \sig_k)^2} \ .
\llabel{130610a} 
\ee


For static perturbations ($\sig_k = 0$) the deformation of the planet is maximal (Eq.\,\ref{130923c}), but the phase lag $\phid_k$ is zero, so there is no tidal dissipation. 
All gravity coefficients are given by their equilibrium values (Eqs.~\ref{130104e}$-$\ref{130104g}), and the last term in expression (\ref{130104b}) and expression (\ref{130104d}) become zero. 
Then, the only effect of the deformation is to add some precession to the elliptical orbit.
The same is true for short period perturbations ($\sig_k \gg 1/\taub $),
since we also have $k_2/Q_k \rightarrow 0$ (Fig.\,\ref{Fig1}c). 
However, in this case the equilibrium figure must be corrected using $\ke$ instead of $\kf$ (Eq.\,\ref{131007a}).

\section{Equilibrium configurations}
\llabel{sores}

\subsection{Low frequency regime ($| \taub \sig_k | \ll 1 $ for all $k$)}

\llabel{lfrjl}
When $\taub\sig_k \ll 1 $  we have  $1 + \taub^2\sig_k^2\approx 1$, and the expression of the secular evolution of the rotation rate (\ref{ddotthetaM}) simplifies. With $\sig_k= 2 \om-kn$, and using  appendix~\ref{apenB} we obtain the following expression for the secular tidal torque
 
\be
\torque = - \frac{3 G \M^2 R^5}{C a^6} \kf \, n \Delta \tau \left(  f_1(e) \frac{\om}{n} - f_2(e) \right) \ , \llabel{131119b}
\ee
where $\Delta \tau = \taub -\taua = \tauv$ is a constant time-lag between the perturbation and the maximal deformation, that is equivalent to the fluid or ``viscous'' relaxation time (Eq.\,\ref{131021e}),
\be
f_1(e) = \frac{1 + 3e^2 + \frac{3}{8}e^4}{(1-e^2)^{9/2}} \ , \llabel{090514n}
\ee
and
\be
f_2(e) = \frac{1 + \frac{15}{2}e^2 + \frac{45}{8}e^4 + \frac{5}{16} e^6}{(1-e^2)^{6}} \ . \llabel{090514o}
\ee
Thus, for a given eccentricity, there is only one possible equilibrium configuration for the rotation rate, obtained when $\torque = 0$, 
\be
\om_\eq / n = f_2(e) / f_1(e) \ , \llabel{131029a}
\ee
which is often known as the {\it pseudo-synchronization}.
The low frequency regime is widely described in the literature, and it is usually known as the ``viscous'' or constant time-lag approximation \citep[e.g.,][]{Singer_1968, Alexander_1973, Mignard_1979}.

\subsection{High frequency regime ($\taub n \gtrsim 1 $)}
\llabel{hfr}

In this case, the situation for equilibrium points ($\torque = 0$) is
more complex than in the low frequency regime as the expression
(\ref{ddotthetaM}) may have many zeros. The situation is very similar
to that discussed for the Andrade rheology model \citep[e.g.,][]{Makarov_Efroimsky_2013}. 
Indeed, the torque (Eq.\,\ref{ddotthetaM}) is a sum 
\be
\torque = \KK \sum_{k=-\infty}^\infty \left(X_k^{-3,2}\right)^2 \torque_k^{(M)}
\llabel{torque}
\ee
with components of the form
\be
\torque_k^{(M)} = \frac{-\taub \sig_k}{1+\tau^2\sig_k^2}\ .
\ee
With Andrade rheology, the torque has an equivalent expression, where all $T_k^{(M)}$ are replaced by $T_k^{(A)}$ defined as \citep{Efroimsky_2012}
\be
\torque_k^{(A)} = \frac{{\cal I}}{{\cal R}^2 + {\cal I}^2}
\mathrm{sign}(\taub\sig_k)
\ee
with
\begin{eqnarray}
{\cal R} &=& 1 + \frac{1}{|\taub \sig_k|^{\alpha}}
\left(\frac{\taua}{\taub}\right)^{1-\alpha}
\cos\left(\frac{\alpha\pi}{2}\right)\,\Gamma(\alpha+1) \ , \crm
{\cal I} &=& -\frac{1}{|\taub\sig_k|} -
\frac{1}{|\taub\sig_k|^{\alpha}} \left(\frac{\taua}{\taub}\right)^{1-\alpha}
\sin\left(\frac{\alpha\pi}{2}\right)\,\Gamma(\alpha+1) \ ,
\end{eqnarray}
where $\alpha$ is an empirical adjustable parameter whose value depends on the material.
\begin{figure}
\centering
\includegraphics*[width=\columnwidth]{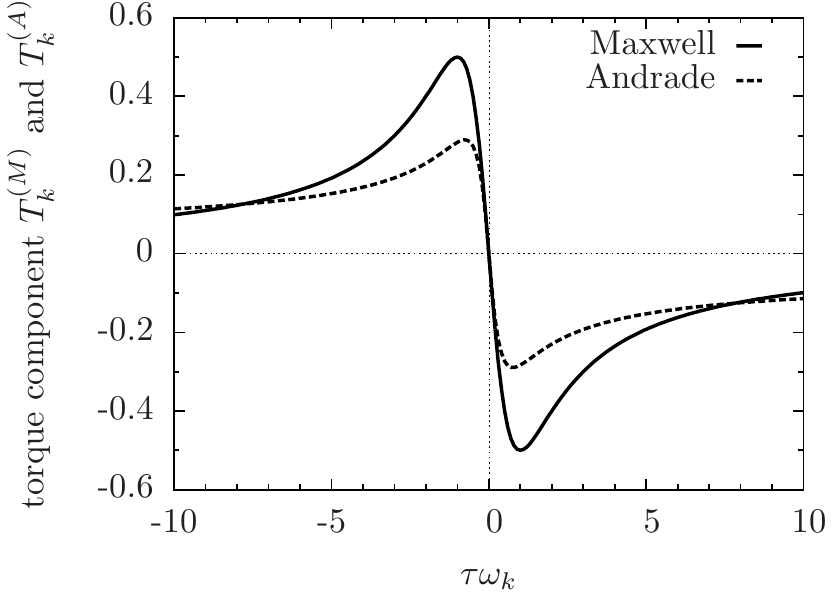}
\caption{Comparison between Maxwell and Andrade secular tidal torque
components $\torque_k$. Parameters of Andrade rheology are $\alpha=0.2$ and $\taua/\taub=0.73$. \llabel{Fig1b} }
\end{figure}%
As shown in Fig.~\ref{Fig1b}, both models have qualitatively similar
asymptotic behaviors. Thus, as explained in
\citet{Makarov_Efroimsky_2013}, several spin-orbit resonant
configurations are at equilibrium $\torque = 0$.

\begin{figure*}
\centering
\includegraphics*[width=\textwidth]{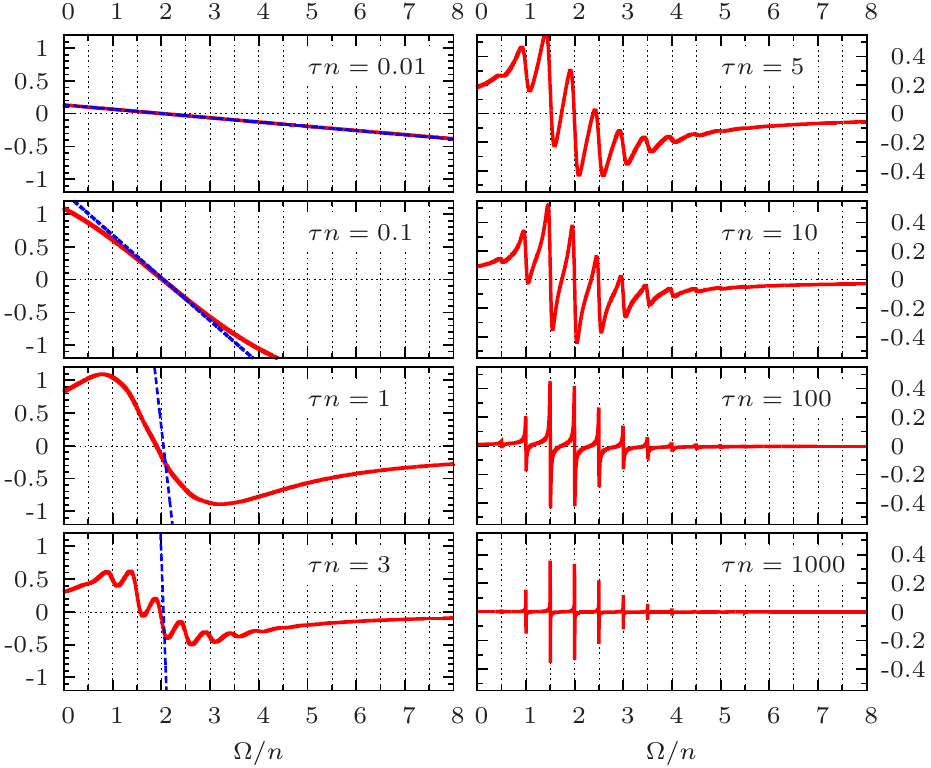}
\caption{Normalized torque $\sum_k \,(X_k^{-3,2})^2 T_k^{(M)}$ for different
values of $\taub n$ and computed at a fixed eccentricity $e=0.4$. The blue dashed line gives the tidal torque corresponding to the linear model (Eq.\,\ref{131119b}), whose equilibrium corresponds to the {\it pseudo-synchronization} (Eq.\,\ref{131029a}).
Hansen coefficients have been obtained by FFT as described in
appendix~\ref{appendix.Hansen}.
\llabel{Fig1c} }
\end{figure*}%

In figure~\ref{Fig1c}, we show the evolution of the secular tidal
torque as $\taub n$ increases from 1 to 1000 for a given eccentricity
$e=0.4$. As long as $\taub n \lesssim 1$, there is only one equilibrium
rotation state $T=0$ which corresponds to the {\it pseudo-synchronous}
state. However, as $\taub n$ increases beyond $\sim 10$, around each spin-orbit
resonance the torque is dominated by a single component $T_k^{(M)}$ with
amplitude $(X_k^{-3,2})^2/2$. This leads to many equilibria, but only rotations simultaneously satisfying $T=0$ and $dT/d\om < 0$ are
stable. As a result, only equilibria
close to spin-orbit resonances are stable
\citep{Makarov_Efroimsky_2013, Storch_Lai_2014}. 
Moreover, not all spin-orbit configurations are equilibria, since the torque also has to cross the
$x$-axis. This condition is satisfied around a $p = k_0/2$ spin-orbit
resonance if
\be
{\left[X_{2 p}^{-3,2} (e)\right]^2} \geq 2 \left| \sum_{k=-\infty}^\infty \left[X_k^{-3,2} (e) \right]^2 \frac{\taub\sig_k }{1+\taub^2\sig_k^2} \right| \ .
\llabel{131105m}
\ee
In particular, for the spin-orbit resonance $p=3/2$ (the last before the synchronization), by truncating the series for terms equal or higher than $e^2$, we approximately get stability as long as
\be
e \gtrsim \frac{2}{7}\sqrt{\frac{2}{\taub n}}\ .
\llabel{131118c}
\ee
This last expression can also be used to quickly determine which $\taub$
values allow non-synchronous rotation for a given eccentricity. In
Figure~\ref{Fig2} we plot the critical eccentricities for some
spin-orbit resonances as a function of the product $\taub n$ given by expressions (\ref{131105m}) and (\ref{131118c}). We conclude that as the relaxation
time $\taub$ increases, the spin-orbit equilibria are only broken for lower values of
the eccentricity, i.e., spin-orbit resonances become more stable.

\begin{figure}
\centering
\includegraphics*[width=\columnwidth]{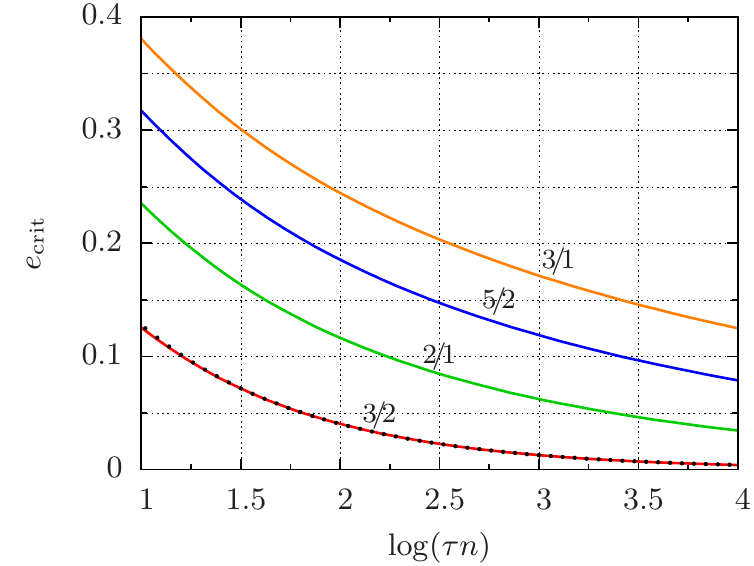}
\caption{Critical eccentricities for keeping the rotation in a spin-orbit resonance as a function of $\taub n$ (Eq.\,\ref{131105m}). Each color corresponds to a different spin-orbit resonance. The dashed line shows the approximation given by expression (\ref{131118c}). Spin-orbit resonances become more stable for higher values of $\taub$. \llabel{Fig2} }
\end{figure}


\section{Libration}
\llabel{libsec}

Most works on tidal dissipation usually assume the zero order solution in $\theta$, the one in the form $\dot \theta = \om = const.$ (section~\ref{tidalevol}).
However, although the tidal drift on the rotation rate can be neglected over one orbital period, near spin-orbit resonances the rotation may present some libration around the equilibrium configuration.
Therefore, we now compute the effect of a low amplitude libration $\theta_1$ such that
\be
\theta = \theta_0 + \theta_1
\ee
with $ \dot \theta_0 = \om = const.$, and
\be
\theta_1 = \Re\left(\rho \ep^{\ii \llambda t} \right)
         = \frac{1}{2}\left(\rho \ep^{\ii \llambda t} + \conj{\rho} \ep^{-\ii \llambda t}\right)\ .
\llabel{eqtheta}
\ee
By assumption, the libration amplitude is small: $|{\rho}| \ll 2\pi$; and the frequency $\llambda$ is secular. We suppose also that there is an integer $k_0$, for which there is a resonant relation
\be
|\sig_{k_0}| = |2\om - k_0 n| \ll \llambda\ ,
\ee
while for all $k\neq k_0$, $|\sig_k| \gg \llambda$.
In this case, the expression of $\Ze$ (Eq.\,\ref{eqzeb}) developped up to the
first order in $|\rho|$ reads
\be
\Ze = \sum_{k=-\infty}^{+\infty} \beta_k \, (1 + 2\ii\theta_1)
\ep^{\ii\sig_kt}\ .
\ee
Thus, from section~\ref{gensol}, we obtain
\begin{eqnarray}
\Z = \sum_{k=-\infty}^{+\infty}
\beta_k \, \bigg(\frac{1+\ii\taua\sig_k}{1+\ii\taub\sig_k}
+\ii\frac{1+\ii\taua(\sig_k+\llambda)}{1+\ii\taub(\sig_k+\llambda)}
\rho\ep^{\ii\llambda t}
\crm 
+\ii\frac{1+\ii\taua(\sig_k-\llambda)}{1+\ii\taub(\sig_k-\llambda)}
\conj{\rho}\ep^{-\ii\llambda t}
\bigg) \ep^{\ii \sig_k t}\ ,
\llabel{eqzzR}
\end{eqnarray}
and 
\begin{eqnarray}
\Delta\Z =  - \left(1- \frac{\taua}{\taub} \right) 
\sum_{k=-\infty}^{+\infty}  \beta_k \, \bigg(
\frac{\ii \taub \sig_k}{1 + \ii \taub \sig_k}
-\frac{\taub(\sig_k+\llambda)}{1+\ii\taub(\sig_k+\llambda)}
\rho\ep^{\ii\llambda t}
\crm
-\frac{\taub(\sig_k-\llambda)}{1+\ii\taub(\sig_k-\llambda)}
\conj{\rho}\ep^{-\ii\llambda t}
\bigg)
\ep^{\ii\sig_kt} \ .
\end{eqnarray}
Since $(\sig_k-\sig_l) t = (k-l)M$, from expression (\ref{ddottheta}) for $\ddot \theta$, we obtain at first order in $\rho$
\be
\ddot \theta = - \KK  \sum_{k=-\infty}^{+\infty} \sum_{l=-\infty}^{+\infty} X_k^{-3,2} X_l^{-3,2} 
\Im\left(z_k \ep^{\ii(k-l)M}\right)\ ,
\llabel{ddotthetacR}
\ee
with
\begin{eqnarray}
z_k = \frac{\ii \taub \sig_k}{1 + \ii \taub \sig_k}
+\left(\frac{\taub\sig_k}{1+\ii\taub\sig_k}
-\frac{\taub(\sig_k+\llambda)}{1+\ii\taub(\sig_k+\llambda)}\right)\rho\ep^{\ii\llambda t}
\crm
+\left(\frac{\taub\sig_k}{1+\ii\taub\sig_k}
-\frac{\taub(\sig_k-\llambda)}{1+\ii\taub(\sig_k-\llambda)}\right)\conj{\rho}\ep^{-\ii\llambda t}
\ .
\llabel{eqzk}
\end{eqnarray}
We now average over the mean anomaly $M$, but keeping the secular
libration $\llambda t$ unaveraged, that is
\be
\langle \ddot \theta \rangle_M = - \KK \sum_{k=-\infty}^{+\infty} \left(X_k^{-3,2}\right)^2 
\Im(z_k)\ .
\llabel{ddotthetacRM}
\ee
\subsection{Low libration frequency regime ($\taub \llambda \ll 1$)}
When $\taub \llambda \ll 1$, the expression (\ref{eqzk}) can be
approximated by
\be
z_k \approx \frac{\ii\taub\sig_k}{1+\ii\taub\sig_k}
\ee
which is exactly the expression without libration. The torque and
the evolution of the system are thus identical to those described in
section~\ref{tidalevol}.

\subsection{High libration frequency regime ($ \taub \llambda \gg 1$)}
Within this regime, it is necessary to distinguish the resonant term
from the others. For $k\neq k_0$, we have $|\sig_k|\gg\llambda$, and thus
expression (\ref{eqzk}) is again
\be
z_k \approx \frac{\ii\taub\sig_k}{1+\ii\taub\sig_k}\ .
\llabel{eqzknR}
\ee
Conversely, for $k=k_0$, we have $|\sig_{k_0}| \ll \llambda$, and expression
(\ref{eqzk}) becomes
\be
z_{k_0} \approx \frac{\ii\taub\sig_{k_0}}{1+\ii\taub\sig_{k_0}}
+\frac{2\ii\theta_1}{1+\ii\taub\sig_{k_0}}
\ .
\llabel{eqzkR}
\ee
As a result, 
combining expressions (\ref{ddotthetacRM}), (\ref{eqzknR}), and (\ref{eqzkR}), we get
\be
\langle\ddot\theta\rangle_M = \torque - \llambda^2 \theta_1\ ,
\llabel{eqddtt1}
\ee
where $\torque$ is the non-resonant torque (Eq.\,\ref{ddotthetaM}),
and
\be
\llambda^2 = 2 \KK \, 
{\left(X_{k_0}^{-3,2}\right)^2}\frac{1}{1+\taub^2\sig_{k_0}^2}\ .
\ee
As in the previous section, the non-resonant torque is responsible for the 
slow variation of $\om$, i.e., $\dot\om = \torque$,
and it remains
\be
\langle\ddot\theta_1\rangle_M + \llambda^2 \theta_1 = 0
\llabel{eqpendule}
\ee
whose solution corresponds to our hypothesis (Eq.\,\ref{eqtheta}).

\section{Application to close-in planets}

\llabel{appexo}

In this section we apply the model described in Sect.\,\ref{TheModel} to
two distinct situations of exoplanets: hot Jupiters and hot super-Earths (Table~\ref{T1}).
For that purpose, we numerically integrate the set of equations (\ref{130104b}$-$\ref{130107g}) for the deformation of the planet, that simultaneously account for orbital and spin evolution due to the rotational flattening and to tides raised by the parent star.

In Table~\ref{T1} we list the observational parameters for the planets considered in the present study.
All these planets were
detected through the transiting method combined with radial velocity measurements. 
Therefore, all orbital parameters are known with good precision, as well as the planetary masses and radius.
Thus, the only unknown parameters in the model are the Love numbers ($\ke$ and $\kf$) and the relaxation times ($\taua$ and $\taub$).

The fluid Love numbers depend on the internal differentiation of the bodies\footnotemark[\value{footnote}].
In the Solar System we have $\kf \approx 0.9$ for the terrestrial planets, and $0.3 <\kf < 0.5 $ for the gaseous planets \citep{Yoder_1995cnt}.
The elastic Love numbers were measured with accuracy only for the Earth and Mars, respectively, $\ke \approx 0.3$ \citep{Mathews_etal_1995}, and $\ke \approx 0.15$ \citep{Konopliv_etal_2006}.
For the gaseous planets only the ratio $k_2/Q$ is known for Jupiter and Saturn, respectively, $k_2/Q \approx 1.1 \times 10^{-5}$ \citep{Lainey_etal_2009}, and $k_2/Q \approx 2.3 \times 10^{-4}$ \citep{Lainey_etal_2012}.
The relaxation times are totally unknown for both terrestrial and gaseous planets, but if one is able to estimate $Q$, then $\taub$ and $\taua$ can be estimated through expressions (\ref{131007a}) and (\ref{130610a}).


 \begin{table}
 \caption{Observational data for the planets HD\,80606\,b \citep{Hebrard_etal_2010b}, HAT-P-34\,b \citep{Bakos_etal_2012}, 55\,Cnc\,e \citep{Gillon_etal_2012b}, and Kepler-78\,b \citep{Howard_etal_2013}.
 \label{T1} }
 \begin{center}
 \begin{tabular}{l | c c c c}
 \hline\hline
\bf Param. &  \bf HD\,80606\,b & \bf HAT-P-34\,b & \bf 55\,Cnc\,e & \bf Kepler-78\,b \\ 
 \hline
 $\M$ [$M_\odot$] & 1.01 &1.392 & 0.905 & 0.83 \\
 $\m$ [$M_\oplus$] & 1297. & 1058. & 8.63 & 1.69 \\
 $R$ [$R_\oplus$] & 10.75 & 13.12 & 2.21 & 1.20 \\
 $P$ [day] & 111.437 & 5.4527 & 0.73654 & 0.35501 \\
 $a$ [au] &0.455 & $0.0677$&  0.0156 & 0.00922 \\
 $e$ & 0.9330 & 0.441 & 0.057 & 0 (fixed) \\
  \hline
 \end{tabular}
 \end{center}
\end{table}

\subsection{Hot Jupiters}

The elastic response of giant gaseous planets is unknown.
Since most of the mass is in their huge atmospheres, for simplicity we assume that only the fluid deformation is important, that is, we adopt a purely viscous rheology.
We note, however, that the core of these planets also experiences tidal effects, that in some cases can be as strong as those present in the convective envelope \citep{Remus_etal_2012a, Guenel_etal_2014}.
In addition, other tidal mechanisms, like the excitation of inertial waves, are expected to take place in fluid planets, that also enhance the tidal dissipation \citep[e.g.,][]{Ogilvie_Lin_2004, Favier_etal_2014}.

In a viscous rheology model, we have $\taua=0$ and $\ke=0$.
This model is equivalent to the one proposed by \citet{Ferraz-Mello_2013}, who adopts
 $Q \sim 10^5$ and $\nue \sim 10^{11}$~Pa\,s, and then obtains from expression (\ref{130524b}) that $\taub \sim 1$~s.
This timescale is intimately connected with the viscous time-lag $\Delta \tau = \tauv$ between the perturbation and high-tide observed ($ \tauv \sim \delta/\sig$, Eq.\,\ref{131004z}).
Nevertheless, 
\citet{Zahn_1977} has shown that for stars possessing a convective envelope, most of the energy flux is transported by turbulent convection, leading to a strong delay of the
equilibrium tide, a ``convective viscous time'' $\tauc \sim 1$~yr, that can be related with the viscous time through \citep[][Eq.\,4.2]{Zahn_1977}
\be
\tauc \approx \frac{R^3}{G \m} \frac{1}{\tauv} \llabel{131014a} \ .
\ee
The ``convective time'' can also be considered for gaseous planets (that is, adopting $\taub = \tauc$), since they can admit regions of turbulent convection \citep[e.g.,][]{Guillot_1999, Ogilvie_Lin_2004}. 
For planets, the effective viscosity of turbulent convection falls rapidly as the oscillation frequency is increased, and in the limit of high tidal frequencies, there is a viscoelastic response to deformation \citep{Ogilvie_Lesur_2012}.
However, for low tidal frequencies (for instance, a spin-orbit commensurability) some different results should be expected.
Therefore, the full deformation of gaseous planets may be on the order of a few years or even decades \citep[for a review see][]{Socrates_etal_2013}.

Hot Jupiters are close-in Jupiter-mass planets with orbital periods less than 10~days.
The in-situ formation of these planets is unlikely, because of the insufficient disk mass close to the star. 
The Solar System formation theories suggest that giant planets preferentially form close to the ice-line where water is condensed into ice and the necessary building blocks for the formation of planets could be found in large quantities \citep[e.g.,][]{Ida_Lin_2008}.
Their presence at a close-in orbit can be explained by considering core-accretion simultaneously with disk driven migration \citep[e.g.,][]{Lin_etal_1996, Alibert_etal_2005, Mordasini_etal_2009a}.
An alternative explanation is obtained through gravitational interactions with a distant inclined binary companion \citep[e.g.,][]{Wu_Murray_2003, Correia_etal_2011} or planet-planet scattering \citep[e.g][]{Rasio_Ford_1996, Beauge_Nesvorny_2012} combined with tidal effects.
Contrarily to the migration scenario, where the eccentricities are quickly damped to zero, the gravitational interaction mechanism can also explain why some proto-hot Jupiters are still observed with high eccentricity values \citep[e.g.,][]{Dawson_Murray-Clay_2013}.
Since for circular orbits the only possibility for the spin is synchronous rotation \citep[e.g.,][]{MacDonald_1964,Hut_1980}, in this paper we focus on eccentric planets.
The solution for circular orbits is also obtained at the end of the evolution for initially eccentric planets.

Here, we apply our model to the planets HAT-P-34\,b \citep{Bakos_etal_2012} and HD\,80606\,b \citep{Hebrard_etal_2010b}, both belonging to single-planet systems. 
The main difference between these two planets is their present orbital period, only 5.45~day for HAT-P-34\,b, and about 111~day for HD\,80606\,b (Table~\ref{T1}).
However, due to the angular momentum conservation, when the eccentricity becomes fully damped, the final semi-major axis and orbital period are constrained \citep[e.g.,][]{Correia_2009}
\be
a_f = a (1-e^2) \quad \Leftrightarrow \quad P_f = P (1-e^2)^{3/2} \ , \llabel{131014b}
\ee
This gives 3.94~day and 5.19~day for HAT-P-34\,b and HD\,80606\,b, respectively, that is, both planets will evolve to ``regular'' hot Jupiter configurations.
For HAT-P-34\,b we begin our simulations with $P = 10.82$~day and $e=0.7$, while for  HD\,80606\,b we start with the present observed values (Table~\ref{T1}).
The initial rotation period is $0.5$~day for both planets, and the planet is assumed perfectly spherical ($J_2 = C_{22} = S_{22} = 0$).
Because the relaxation time is unknown, we adopt different values for $\taub$ ranging from $10^{-5}-10^0$~yr, in order to cover all kinds of behavior. 
For the fluid Love number we use Jupiter's value $\kf = 0.5$ \citep{Yoder_1995cnt}.

In Figure~\ref{Fig3} we show the future evolution of the HD\,80606\,b orbit (top), rotation (middle) and shape (bottom) using different values of the relaxation time.
The shape of the planet is described by its gravity field coefficients.
However, since $C_{22}$ and $S_{22}$ values constantly change for a non-synchronous planet, we plot the prolateness of the planet,
\be
\epsilon = | \Z | = \sqrt{C_{22}^2+S_{22}^2} \ , \llabel{140103a}
\ee
which is equivalent to the value of $C_{22}$ along the instantaneous principal long axis of inertia.
In all situations, the planet attempts to point the instantaneous long axis to the star, increasing its prolateness until it reaches the pericenter of the orbit.
At this stage, the prolateness decreases again, reaching a minima at the apocenter.

\begin{figure*}
\centering
\includegraphics*[width=\textwidth]{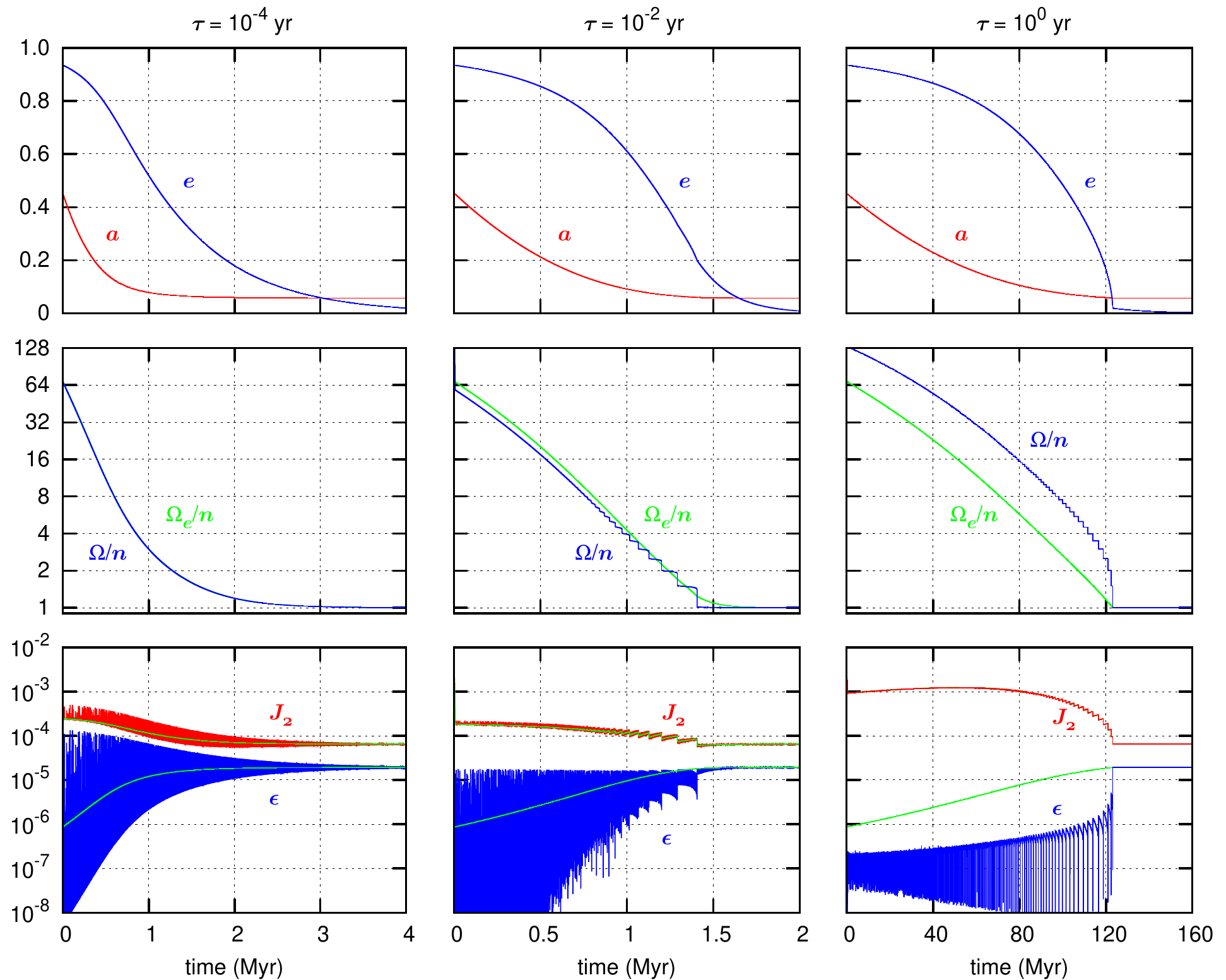}
\caption{Evolution of HD\,80606\,b with time for different $\taub$ values. We plot the semi-major axis (in AU) and the eccentricity ({\it top}), the ratio between the planet rotation rate and the orbital mean motion ({\it middle}), and the planet $J_2$ and $\epsilon$ ({\it bottom}). The green line gives the {\it pseudo-synchronization} equilibrium (Eq.\,\ref{131029a}) ({\it middle}), and the average equilibrium values for $J_2$ and $\epsilon$, respectively (Eqs.~\ref{140103b} and \ref{140103c}) ({\it bottom}).  \llabel{Fig3} }
\end{figure*}

For $\taub = 10^{-4}$~yr (Fig.\,\ref{Fig3}, left), the relaxation time is much shorter than the orbital period, so between two consecutive passages at the apocenter, the prolateness of the planet is back to its minimal value.
The only effect on the rotation rate is then to decrease it from its initial value, until it reaches the {\it pseudo-synchronization} equilibrium (Eq.\,\ref{131029a}), since we have $ \taub \sig_k \ll 1$ (section~\ref{lfrjl}).
Because of the conservation of the angular momentum, the eccentricity of the orbit is also progressively damped to zero.
When the eccentricity is close to zero, $\om / n \approx 1 + 6e^2$ (Eq.\,\ref{131029a}), so the equilibrium rotation tends to the synchronous rotation.
At this stage the planet is able to acquire a permanent prolateness, always pointing the long axis to the star.

For $\taub = 10^{-2}$~yr (Fig.\,\ref{Fig3}, center), the relaxation time is also shorter than the orbital period during the initial stages of the spin evolution, so the rotation behaves similarly to the previous case ($\taub = 10^{-4}$~yr).
However, as the orbit shrinks, the relaxation time and the orbital period become equivalent $\taub n \sim 1 $.
Then, when the planet is back at the pericenter, its prolateness is still excited by the previous pericenter passage.
As a consequence, the planet increases its global prolateness at each pericenter passage.
This effect is maximized near a spin-orbit commensurability, 
since the rotation rate is close to a harmonic of the orbital mean motion, $\om / n = p = k_0/2$, and thus there exist one tidal frequency $\sig_{2p}$ that is near zero (section~\ref{hfr}).
The gravitational torque exerted by the star on the prolate planet dominates the tidal toque and tends to keep the rotation near the spin-orbit commensurability (Fig.\,\ref{Fig1c}).
Therefore, in this regime the rotation rate closely follows the spin-orbit equilibria.

Finally, for $\taub = 1$~yr (Fig.\,\ref{Fig3}, right), the relaxation time is permanently larger than the orbital period. 
Therefore, the prolateness of the planet is continuously excited by a nearby spin-orbit resonance, as explained for the final stages in the previous case.
The spin remains trapped in a spin-orbit equilibrium until the eccentricity decreases below a given threshold (Fig.\,\ref{Fig2}).
Then, the spin decreases one more step until it is captured in the following low-order spin-orbit resonance.
The process continues until the synchronous equilibrium is reached for small values of the eccentricity, since this is the only spin-orbit resonance that is always stable (Eq.\,\ref{131105m}).

For lower and higher $\taub$ values the system asymptotically behaves as in the $\taub = 10^{-5}$~yr and $10^0$~yr cases, respectively, the major change is in the evolution timescale, that is longer.

\subsection{Evolution versus eccentricity}

In Figure~\ref{Fig3}, the evolution timescale is different for each $\taub$ value, since tidal dissipation is controled by this parameter (Fig.\,\ref{Fig1}).
However, because of the conservation of the angular momentum (Eq.\,\ref{131014b}), the final semi-major axis is always the same.
In addition, there is a perfect correspondence between the semi-major axis and the eccentricity (Fig.\,\ref{Fig4}), that is, the orbital evolution is always the same, only the timescale changes.
Therefore, in order to better compare the behavior of the spin with different $\taub$ values, instead of using the time, it is preferable to use the semi-major axis or the eccentricity in the evolution axis.

\begin{figure}
\centering
\includegraphics*[width=\columnwidth]{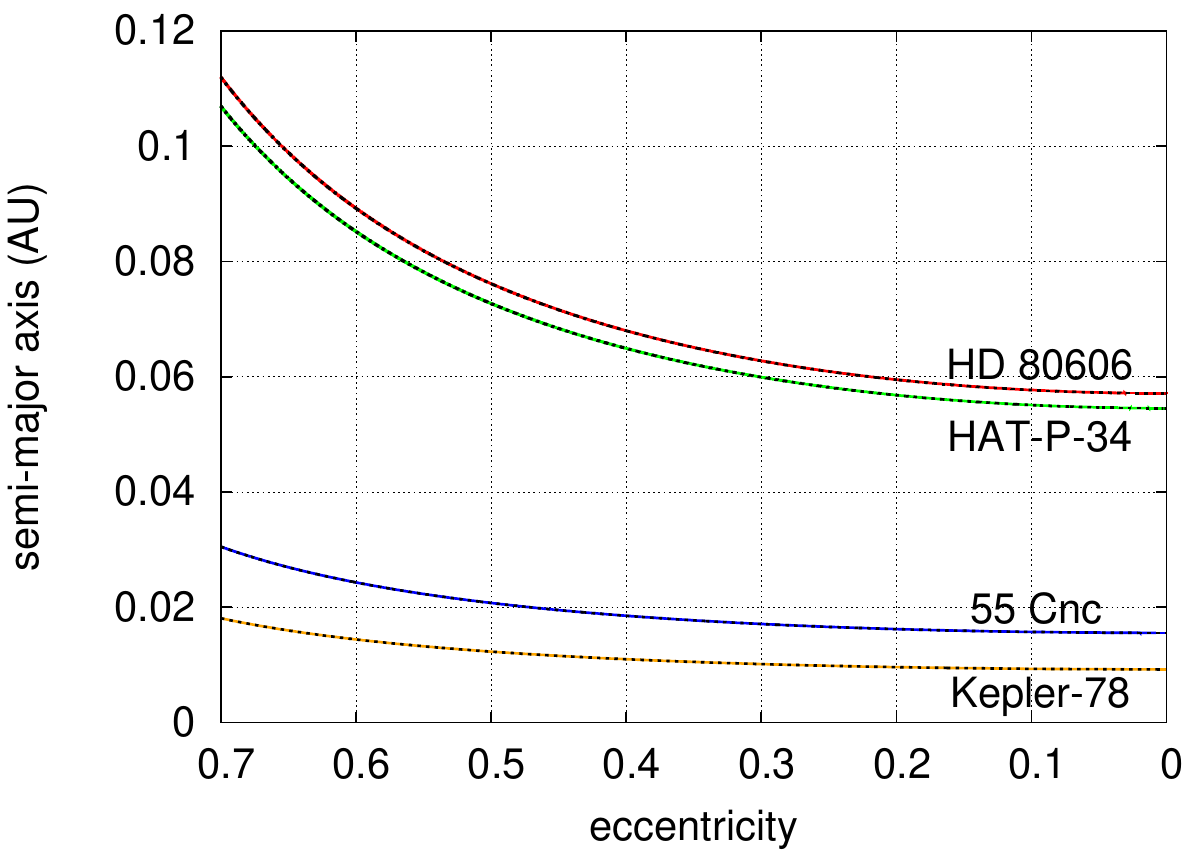}
\caption{Evolution of the semi-major axis as a function of the eccentricity for different systems and $\taub$ values. The dotted lines correspond to the theoretical prediction given by expression (\ref{131014b}). Solid lines correspond to the orbital evolution obtained using numerical simulations with different $\taub$ values.  For each system, the simulations exactly match the theoretical prediction.
 \llabel{Fig4} }
\end{figure}

In Figure~\ref{Fig5}, we plot the evolution of the spin of HAT-P-34\,b and HD\,80606\,b as a function of the eccentricity for different $\taub$ values. 
The initial rotation period is arbitrarily set at $0.5$~day for both planets in all simulations.
This value is not critical, because for all $\taub$ values the tidal torque on the spin is so strong that the rotation quickly evolves into an equilibrium configuration.
Indeed, we observe that the spin evolution of both planets is identical, although the initial eccentricity of HAT-P-34 is 0.7, while the initial eccentricity of HD\,80606 is 0.933.
We can also clearly observe the two tidal regimes described in section~\ref{sores}.
For low eccentricity values, the orbital period of both planets is around $10^{-2}$~yr.
Thus, for $\taub \ll 10^{-2} $\,yr, the spin closely follows the {\it pseudo-synchronization} equilibrium (Eq.\,\ref{131029a}), since $ \taub n \ll 1$. 
For $\taub \sim 10^{-2} $\,yr, we observe the transition between regimes, while
for $\taub \gg 10^{-2} $\,yr, the spin is captured in half-integer spin-orbit resonances, since $  \taub n \gg 1$.
These resonances become destabilized as the eccentricity decays, in agreement with expression (\ref{131105m}). The main consequence of these metastable equilibria is to allow faster rotation rates for an identical eccentricity.

\begin{figure}
\centering
\includegraphics*[width=\columnwidth]{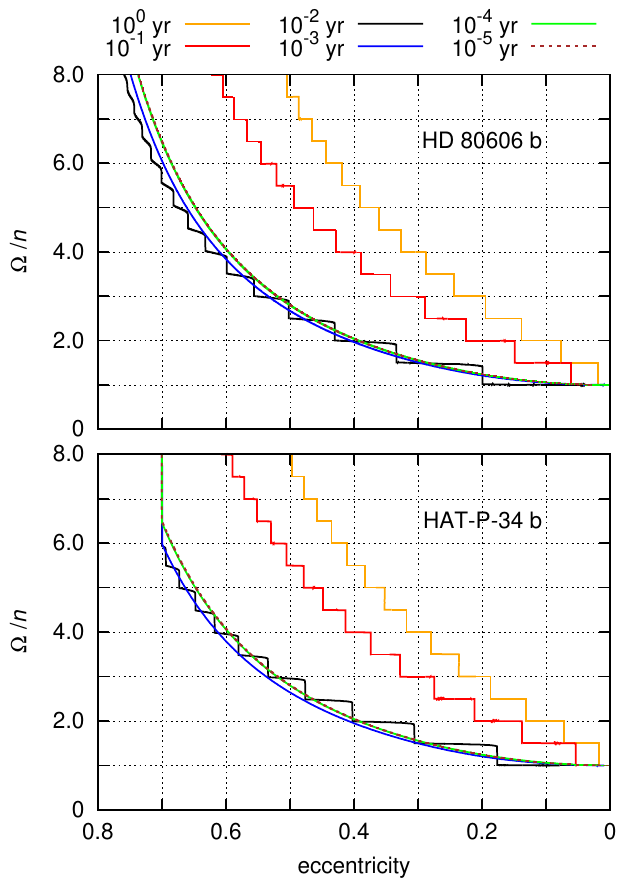}
\caption{Evolution of the rotation rate as a function of the eccentricity for different $\taub$ values ($10^{-5} \le \taub \le 10^{0}$~yr). The initial rotation period is $0.5$~day for both planets, but this value is not critical as the spin quickly evolves to an equilibrium configuration.
For $\taub < 10^{-3} $\,yr, the spin closely follows the {\it pseudo-synchronization} equilibrium (Eq.\,\ref{131029a}). For $\taub > 10^{-2} $\,yr, the spin is captured in half-integer spin-orbit resonances that become destabilized as the eccentricity decays, in agreement with expression (\ref{131105m}). \llabel{Fig5} }
\end{figure}

In Figure~\ref{Fig6}, we plot the evolution of the shape ($J_2$ and $\epsilon$) of HD\,80606\,b as a function of the eccentricity.
To better understand the different behaviors, we also plot the average of the equilibrium shape over one orbital period (Eqs.~\ref{130104e}$-$\ref{130104g})
\be
\left\langle J_2 \right\rangle_M  = \kf \left[ \frac{\om^2 R^3}{3 G \m} + \frac{1}{2} \frac{\M}{\m} \left(\frac{R}{a} \right)^3 (1-e^2)^{-3/2} \right]  \llabel{140103b} \ ,
\ee
\be
\left\langle \epsilon \right\rangle_M =  \frac{\kf}{4} \frac{\M}{\m} \left(\frac{R}{a}\right)^3 (1-e^2)^{-3/2} \ , \llabel{140103c} 
\ee
as well as the equilibrium shape for a planet trapped in a specific spin-orbit resonance $p=k_0/2$ (Eq.\,\ref{eqzeb})
\be
\left\langle \epsilon_p \right\rangle_M =  \beta_{2p} = \frac{\kf}{4} \frac{\M}{\m} \left(\frac{R}{a}\right)^3 X_{2p}^{-3,2} (e)  \ . \llabel{140103d} 
\ee

\begin{figure}
\centering
\includegraphics*[width=8.6truecm]{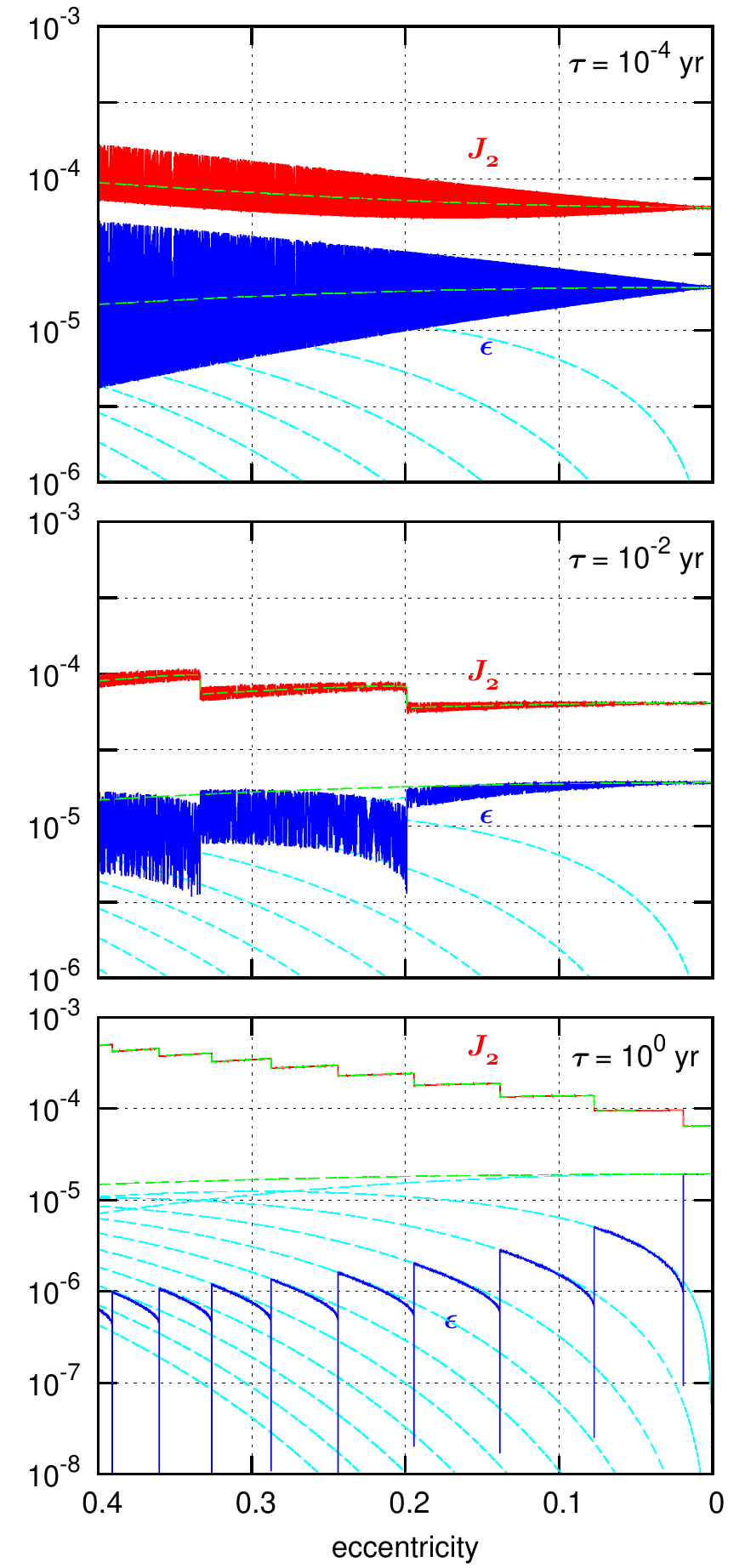}
\caption{Evolution of $J_2$ (in red) and prolateness $\epsilon$ (in blue) of HD\,80606\,b as a function of the eccentricity for different $\taub$ values. 
The green dashed line shows the average equilibrium value of these two quantities over one orbital period (Eqs.~\ref{140103b}, \ref{140103c}). The cyan dashed line shows the average equilibrium value of the prolateness for a specific spin-orbit resonance (Eq.\,\ref{140103d}).  \llabel{Fig6} }
\end{figure}

For $\taub \ll 10^{-2} $\,yr, $J_2$ and $\epsilon$ present large oscillations around the average equilibrium shape (Eqs.\,\ref{140103b} and \,\ref{140103c}).
The maximal value corresponds to the equilibrium shape at the pericenter, while the minimal value corresponds to the equilibrium shape at the apocenter, obtained by replacing $r=a (1-e)$ and $r=a (1+e)$ in expressions (\ref{130104e}) to (\ref{130104g}), respectively.
In this regime the relaxation time is much shorter than the orbital period, so the figure of the planet is almost equal to the instantaneous equilibrium figure.

For $\taub \gg 10^{-2} $\,yr, the relaxation time is much longer than the orbital period, so it is possible to temporary lock the spin in resonance.
Therefore, the $J_2$ of the planet closely follows the average equilibrium value (Eq.\,\ref{140103b}), although sudden variations are observed, corresponding to the transition between two spin-orbit resonances.
The prolateness of the planet also follows the equilibrium value for each spin-orbit resonance (Eq.\,\ref{140103d}).
The prolateness temporarily decreases with the eccentricity, since $X_{2p}^{-3,2} (e)$ is a decreasing function with $e$ (check Table\,\ref{TabX}).
However, when the critical eccentricity for each resonance is attained (Eq.\,\ref{131105m}), the prolateness $\epsilon$ increases again stabilizing in the following resonance.

For $\taub = 10^{-2} $\,yr, we observe a mix of both behaviors mentioned above.
The prolateness $\epsilon$ still follows the resonant equilibrium values (Eq.\,\ref{140103d}), but now presents significant oscillations around it.
Resonances also become unstable for higher eccentricity values in agreement with Figure~\ref{Fig2}.
Finally, at the end of the evolution ($e=0$), the planet always evolves into the synchronous resonance, acquiring the same $J_2$ and $\epsilon$ value in all scenarios.

\subsection{Super-Earths}

In general, a super-Earth is defined exclusively by its mass, independently of its environment being similar to that of the Earth. 
An upper bound of 10 Earth masses is commonly accepted to warrant that the planet is mostly rocky \citep[e.g.,][]{Valencia_etal_2007a}.
Therefore, the rheology of these planets is expected to be similar to the terrestrial planets in the Solar System.
We adopt here the full viscoelastic model described in section~\ref{TheModel},
using the Earth's values for the Love numbers, $\ke = 0.3$ and $\kf = 0.9$ \citep{Yoder_1995cnt}.

For the Earth and Mars, we have $Q=10$ \citep{Dickey_etal_1994} and $Q=80$ \citep{Lainey_etal_2007}, respectively.
We then compute for the Earth $\taub=1.6$~day and $\taua=0.5$~day, and for Mars $\taub=14.7$~day and $\taua=2.5$~day.
However, in the case of the Earth, the present $Q$-factor is dominated by the oceans, the Earth's solid body $Q$ is estimated to be 280 \citep{Ray_etal_2001}, which increases the relaxation time by more than one order of magnitude ($\taub=46$~day and $\taua=15$~day).
Although these values provide a good estimation for the average present dissipation ratios, they appear to be incoherent with the observed deformation of the planets.
Indeed, in the case of the Earth, the surface post-glacial rebound due to the last glaciation about $10^4$~years ago is still going on, suggesting that the Earth's mantle relaxation time is something like $\taub=4400$~year \citep{Turcotte_Schubert_2002}.
Therefore, we conclude that the deformation of terrestrial planets can range from a few days up to thousands of years.

Here we apply our model to the planets 55\,Cnc\,e \citep{Gillon_etal_2012b} and Kepler-78\,b \citep{Howard_etal_2013}. 
The main difference between these two planets is their masses, only 1.7~$M_\oplus$ for Kepler-78\,b, and about 8.6~$M_\oplus$ for 55\,Cnc\,e (Table\,\ref{T1}).
55\,Cnc\,e belongs to a multi-planet system, but in the present study we ignore the gravitational interactions with the remaining companions, as in \citet{Rodriguez_etal_2012}.
Contrary to hot Jupiters from the previous section, the present eccentricities of these two super-Earths are very small, nearly zero.
Therefore, as for HAT-P-34\,b, we begin our simulations with $e=0.7$, so that we can observe some tidal evolution.
The initial orbital periods are obtained through expression (\ref{131014b}), which gives about 0.97 and 2.05~day for Kepler-78\,b and 55\,Cnc\,e, respectively.
Because the relaxation time is unknown and can vary by some orders of magnitude, we adopt different values for $\taub$ ranging from $10^{-2}-10^{1}$~yr. 

The initial rotation period is arbitrarily set at $0.5$~day for both planets in all simulations.
This value is not critical, because for all $\taub$ values the tidal torque on the spin is so strong that the spin quickly evolves into a spin-orbit resonance (Figs.~\ref{Fig7} and \ref{Fig5st}).
Indeed, since the initial orbital period is only $\sim 1$~day, we get $\om \sim n$ from the  beginning of the simulations.
For instance, in the case of 55\,Cnc\,e we have $\om / n \approx 4.1$, that is, the spin starts close to the $4/1$ spin-orbit resonance.
However, the initial planet is assumed perfectly spherical ($J_2 = C_{22} = S_{22} = 0$), so during the initial stages of the evolution (on the order of $\taub$), the spin is chaotic, until the gravity field coefficients acquire some stable deformation.
As a consequence, the spin can be stabilized in spin-orbit resonances different from the $4/1$, depending on the initial phase of the rotation angle $\theta$ (Fig.\,\ref{Fig7}).
Although the initial capture in a spin-orbit resonance is random, the subsequent evolution is very predictable: as the eccentricity decreases, the higher-order resonances become unstable, and the spin successively quits them one by one, until it reaches the synchronous rotation at the end of the evolution.

\begin{figure}
\centering
\includegraphics*[width=8.6truecm]{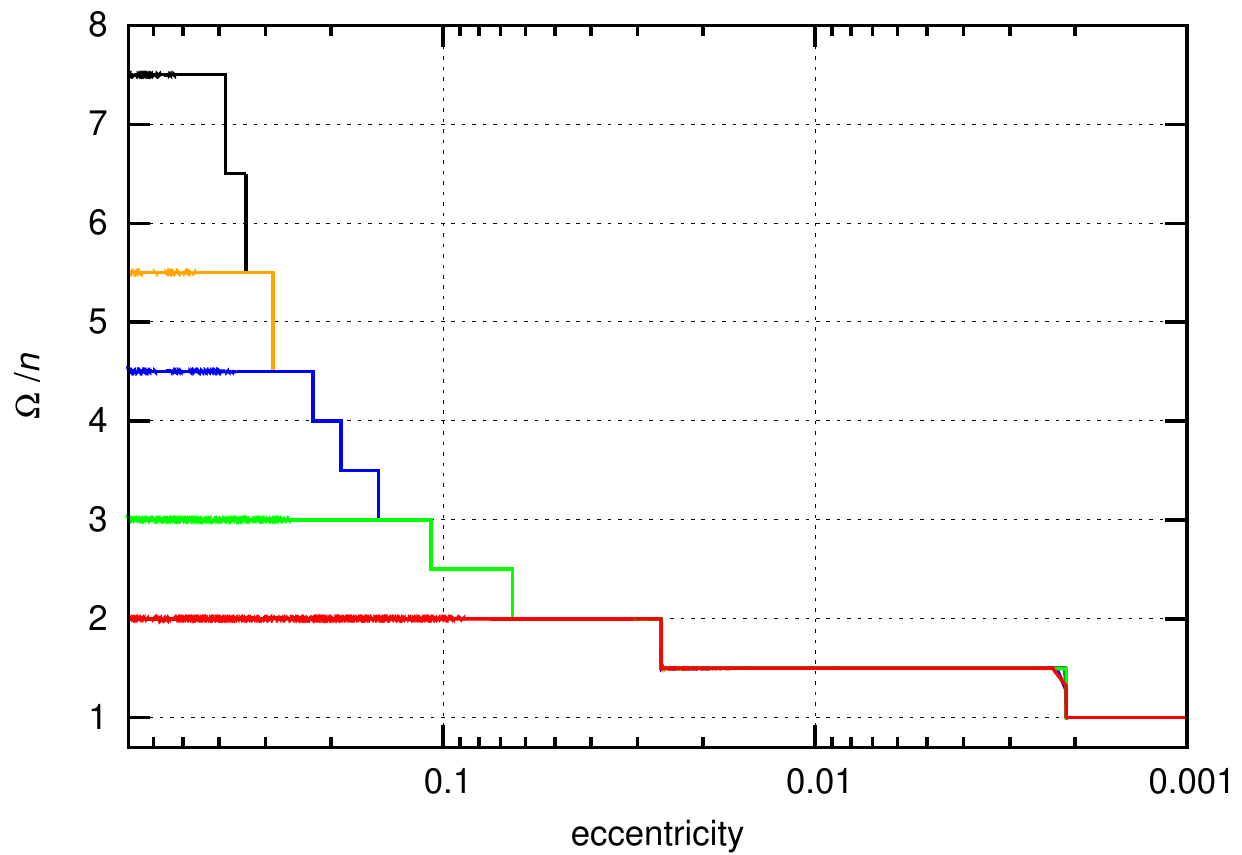}
\caption{Evolution of the rotation of 55\,Cnc\,e as a function of the eccentricity for different initial $\theta$ values and $\taub= 10^{1}$~yr. The initial rotation rate is $\om/n \approx 4.1$, but since the initial shape of the planet is perfectly spherical ($J_2 = C_{22} = S_{22} = 0$), the initial rotation is chaotic and the spin may be trapped in different spin-orbit resonances. As the eccentricity decreases, the higher-order resonances become unstable, and the spin successively quits them, until it reaches the synchronous rotation. \llabel{Fig7} }
\end{figure}

In Figure~\ref{Fig5st}, we plot the evolution of the spin as a function of the eccentricity for different $\taub$ values.
Unlike hot Jupiters, for the considered $\taub$ values (and higher), super-Earths are always found in the high-frequency tidal regime (section~\ref{hfr}). 
Therefore, for both planets the spin is initially captured in a 
spin-orbit resonance, and subsequently evolves to lower-order resonances as the eccentricity decreases.
For higher $\taub$ values, the resonances become more stable, in agreement with expression (\ref{131105m}), so the planet requires a lower eccentricity value before reaching the synchronous rotation.
For planets in multi-body systems like 55\,Cnc\,e, we cannot rule out a non-synchronous rotation at present.
Using the observed value of $e=0.057$ 
in expression (\ref{131118c}), we conclude that 55\,Cnc\,e is only synchronous if $\taub < 10^{-2}$~yr (Fig.\,\ref{Fig5st}).

\begin{figure}
\centering
\includegraphics*[width=\columnwidth]{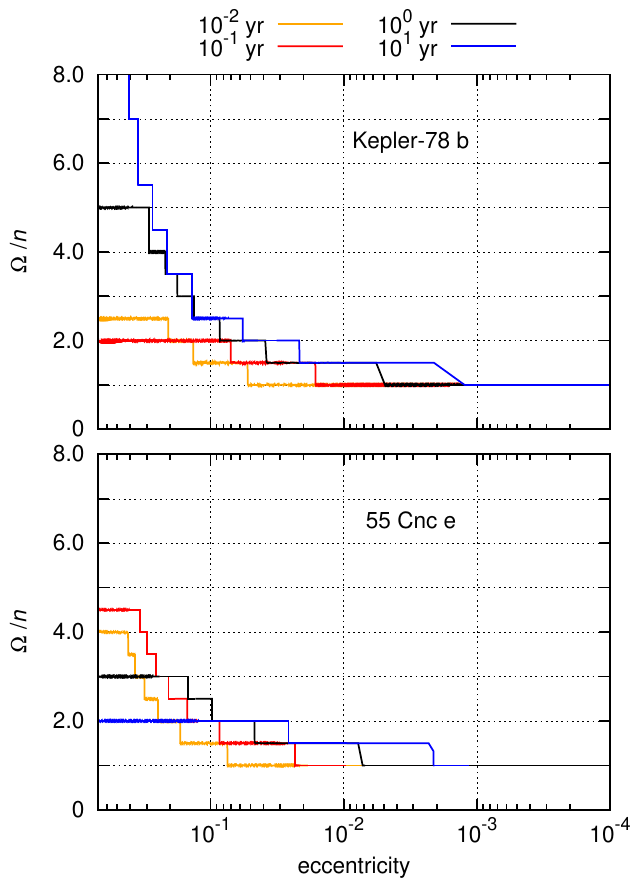}
\caption{Evolution of the rotation rate as a function of the eccentricity for different $\taub$ values ($10^{-2} \le \taub \le 10^{1}$~yr). 
The initial rotation period is $0.5$~day for both planets, but this value is not critical as the spin quickly evolves into a spin-orbit resonance (Fig.\,\ref{Fig7}).
The resonances are destabilized as the eccentricity decays, in agreement with expression (\ref{131105m}). \llabel{Fig5st} }
\end{figure}

In Figure~\ref{Fig6st}, we plot the evolution of the shape ($J_2$ and $\epsilon$) of 55\,Cnc\,e as a function of the eccentricity.
As for hot Jupiters (Fig.\,\ref{Fig6}), we also plot the average of $J_2$ over one orbital period of the equilibrium (Eq.\,\ref{140103b}), as well as the average of $\epsilon$ for a planet trapped in a specific spin-orbit resonance
\be
\left\langle \epsilon_p \right\rangle_M = \frac{\kf}{4} \frac{\M}{\m} \left(\frac{R}{a}\right)^3 \left[ \big(1-\frac{\taua}{\taub} \big) \, X_{2p}^{-3,2} (e) + \frac{\taua}{\taub} (1-e^2)^{-3/2} \right] \ . \llabel{140107a} 
\ee
This expression is different from its analog for hot Jupiters (Eq.\,\ref{140103d}), because here we have an elastic contribution to $\epsilon$, while previously we only considered the viscous contribution ($\taua=0$).
This equilibrium is also independent of the relaxation times, since $\taua/\taub = \ke/\kf$ (Eq.\,\ref{131007a}) (in our simulations $\taua/\taub = 1/3$).

For all $\taub > 10^{-2}$~yr, the $J_2$ of the planet closely follows the average of the equilibrium value (Eq.\,\ref{140103b}), the sudden variations corresponding to the transition between two spin-orbit resonances.
The prolateness of the planet also follows the equilibrium value for each spin-orbit resonance (Eq.\,\ref{140107a}).
However, for all spin-orbit resonances different from the synchronization, the elastic contribution given by the second term in expression (\ref{140107a}) is dominating (in particular for low eccentricity values), since for $p \ne 1$ we have $X_{2p}^{-3,2} (e\rightarrow 0) = 0$ (Table\,\ref{TabX}).
Thus, for all these resonances the observed prolateness converges to the same equilibrium value,
\be
\left\langle \epsilon_p \right\rangle_M \approx 
\frac{\kf}{4} \frac{\taua}{\taub} \frac{\M}{\m} \left(\frac{R}{a}\right)^3 = 
\frac{\ke}{4} \frac{\M}{\m} \left(\frac{R}{a}\right)^3  \ . \llabel{140110a} 
\ee
For the synchronous resonance, we have $X_{2}^{-3,2} (e \rightarrow 0) = 1$ (Tab.\,\ref{TabX}), so both terms in expression (\ref{140107a}) add up to give an equilibrium identical to the previous one (Eq.\,\ref{140110a}), but where $\taua$ and $\ke$ are replaced by $\taub$ and $\kf$, respectively.
This final equilibrium for $\epsilon$ is exactly the same as for hot Jupiters (Eq.\,\ref{140103d}).

\begin{figure}
\centering
\includegraphics*[width=8.6truecm]{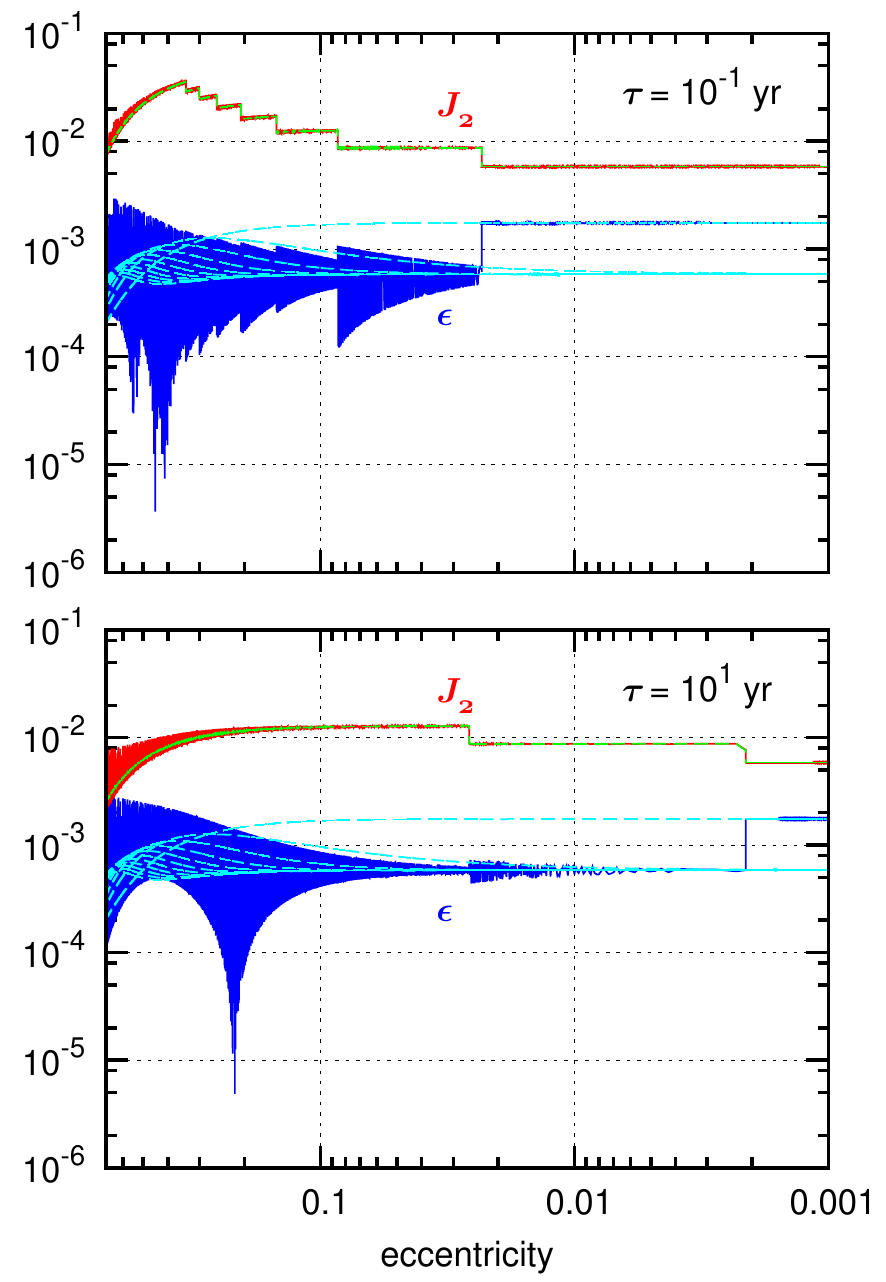}
\caption{Evolution of $J_2$ (in red) and prolateness $\epsilon$ (in blue) of 55\,Cnc\,e as a function of the eccentricity for different $\taub$ values. 
The green dashed line shows the average equilibrium value of $J_2$ over one orbital period (Eq.\,\ref{140103b}). The cyan dashed line shows the average equilibrium value of the prolateness for each spin-orbit resonance (Eq.\,\ref{140107a}).  \llabel{Fig6st} }
\end{figure}

\begin{figure}
\centering
\includegraphics*[width=8.6truecm]{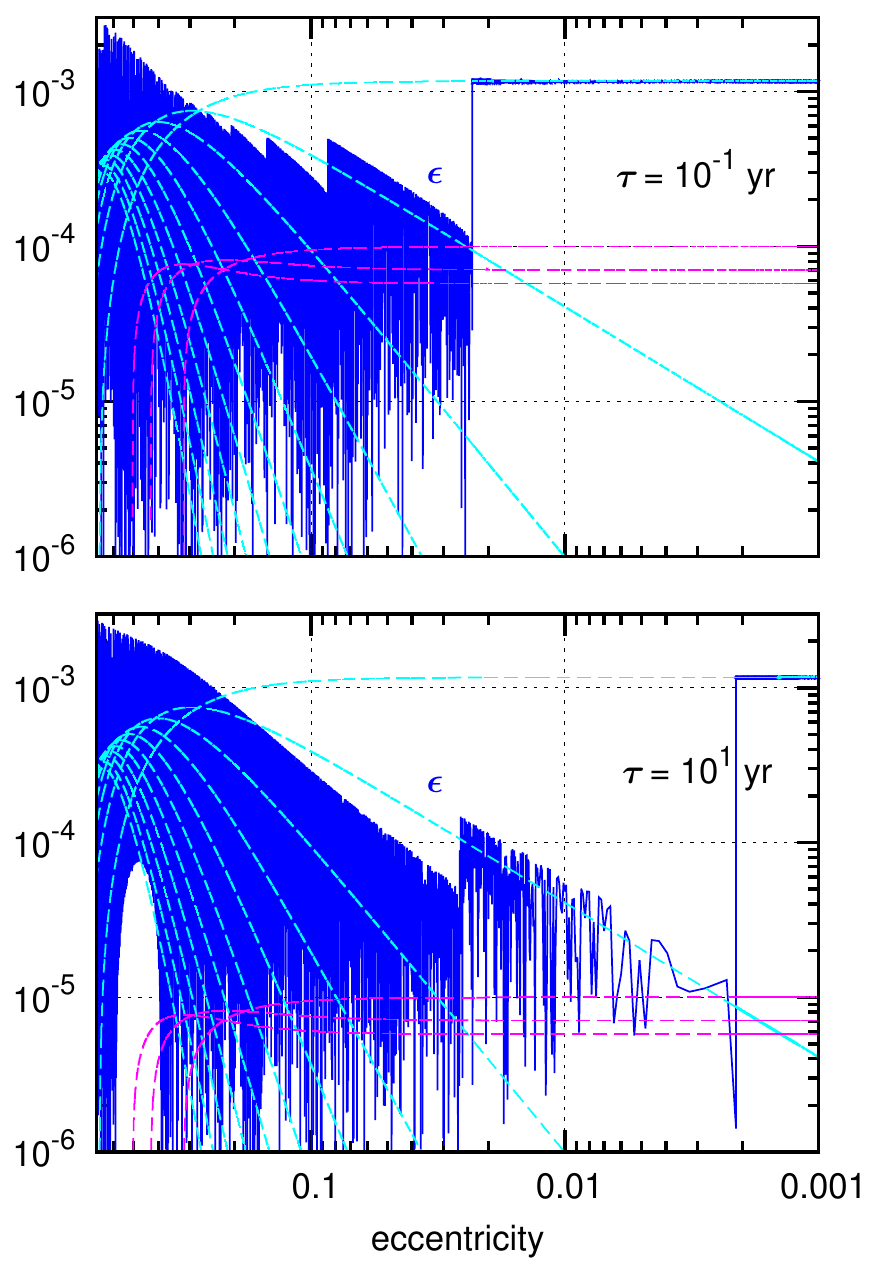}
\caption{Evolution of the ``permanent'' prolateness (in blue) of 55\,Cnc\,e as a function of the eccentricity for different $\taub$ values. 
The cyan dashed lines show the average equilibrium value of the ``permanent'' prolateness for each spin-orbit resonance (first term in Eq.\,\ref{140107a}).
The pink dashed lines give the strength of the tidal torque for each resonance (Eq.\,\ref{ddotthetaM}).  \llabel{Fig6zst} }
\end{figure}

Despite $\epsilon$ has a dominating ``elastic'' contribution, the first term in expression (\ref{140107a}) corresponds to a ``permanent'' deformation.
That is, while the deformation that results from the contribution of the first term is along an axis that is fixed with respect to the body, the deformation resulting from the second term nearly points to the direction of the star. 
As a result, only the ``permanent'' contribution is responsible for the capture in resonance.
In Figure~\ref{Fig6zst} we show the evolution of the ``permanet'' prolateness with the eccentricity, i.e., we show $\epsilon$ subtracted from its ``elastic'' term (Eq.\,\ref{140110a}).
As expected, the ``permanent'' prolateness closely follows the equilibrium cyan dashed lines given by the first term in expression (\ref{140107a}).

In Figure~\ref{Fig6zst}, we additionally plot in pink dashed lines the strength of the tidal torque for each spin-orbit resonance (Eq.\,\ref{ddotthetaM}).
We observe that the equilibrium in each resonance is broken exactly when the cyan and pink dashed lines intercept.
This result is in perfect agreement with the theoretical prediction from expression (\ref{131105m}), and confirms that only the ``permanent'' contribution to $\epsilon$ is  responsible for the capture in resonance.


\section{Conclusions}

In this paper we presented a new approach to the tidal theory. 
Adopting the same strategy as \citet{Darwin_1879a}, we are able to extend the creep equation obtained for an homogeneous incompressible viscous body to also include the elastic rebound.
The deformation law hence obtained (Eq.\,\ref{131021e}) corresponds to the deformation of a Maxwell material.
The deformation is then taken into account by means of the effect of the disturbing potential on the gravity field coefficients of the planet (Eq.\,\ref{130524f}).
This method allows us to easily compute the instantaneous variations in the shape of the planet when subject to a perturbation.

In this study we considered the planar case, where the spin of the planet is orthogonal to its orbit.
The planet is deformed under the effect of tidal perturbations, but also by the changes in its rotation due to the centrifugal potential.
Our model can be easily extended to non-planar configurations (for planets with some obliquity and evolving in inclined orbits), provided that we additionally take into account the deformation of the $C_{21}$ and $S_{21}$ gravity field coefficients in the gravitational potential (Eq.\,\ref{130528a}), as explained in \citet{Correia_Rodriguez_2013}.

There are several advantages in using our model.
First, it works for any kind of perturbation, even for the non-periodic ones (such as chaotic motions or transient events).
Thus, we no longer need to decompose the tidal potential in an infinite sum of harmonics of the tidal frequency \citep[e.g.,][]{Kaula_1964, Mathis_Poncin-Lafitte_2009, Efroimsky_Williams_2009}.
Second, as long as we can neglect terms in $(R/r)^3$, the model is valid for any eccentricity value, we do not need to truncate the equations of motion.
Finally, it simultaneously reproduces the deformation and the dissipation on the planet.
This last point is very important, because it clarifies an important behavior of the tidal torque: its ability to temporarily lock the planet in a spin-orbit resonance. 
Indeed, for relaxation times longer than the orbital period ($\taub n \gg 1$), when the planet is back at the pericenter, its shape is still excited by the previous pericenter passage.
The planet thus increases its global prolateness at each pericenter passage, an effect that
is maximized near a spin-orbit commensurability. 
The gravitational torque exerted by the star on the prolate planet dominates the tidal torque and tends to keep the rotation near the spin-orbit commensurability (Fig.\,\ref{Fig1c}).


Another important outcome of our study is that fluid planets can also present non-synchronous spin-orbit equilibria. The only requirement is that the relaxation time is longer than the orbital period.
Although this seems unlikely for the gaseous envelop, giant planets also contain liquid and solid cores that may take longer to acquire their equilibrium shape \citep[e.g.,][]{Remus_etal_2012a, Auclair-Desrotour_etal_2014}.
In particular, we cannot rule out this possibility for hot Jupiters, since they have very short orbital periods.

For close-in super-Earths (or terrestrial planets), the relaxation time of the mantle is almost certainly longer than the orbital period.
As a consequence, our model shows that the rotation is temporarily locked in several spin-orbit resonances throughout the evolution.
These equilibria are destabilized as the eccentricity decreases, but non-synchronous rotation can be maintained even for extremely low eccentricity values, depending on the relaxation time.
Moreover, if there are other planetary companions, the eccentricity is never exactly zero \citep[e.g.,][]{Laskar_etal_2012}, so the higher-order spin-orbit resonances can be maintained for the entire life of the system.
As a consequence, we expect that in many cases, the rotation of close-in terrestrial planets in multi-planet systems is captured in non-synchronous spin-orbit resonances.


Our model is also able to reproduce the results by \citet{Ferraz-Mello_2013} for a Newtonian creep.
For that, we just need to set $\taub = \gamma^{-1}$ and $\taua=0$, and thus neglect expressions (\ref{130107e}$-$\ref{130107g}).
Actually, 
apart from the elastic deformation of the planet, this simplification still allows us to reproduce the main features described in this paper, like the different regimes of dissipation and the spin-orbit capture.
Therefore, for studies where the deformation of the planet does not matter (i.e., only the dissipation is relevant), we can adopt the above simplification using a modified Love number $\kf'$ instead of $\kf$ (Eq.\,\ref{130610a})
\be
\kf' = \kf (1-\taua/\taub) = \kf - \ke \ . \llabel{140131a}
\ee

\citet{Remus_etal_2012a} have shown that for a planet composed of a homogeneous core surrounded by a fluid envelope of constant density (which can be a rocky planet with a deep non-dissipative ocean, or a giant planet with a solid core), the global response to the perturbations can be modeled by a viscoelastic rheology.
As an illustration, they present the results 
for a Maxwell body.
Thus, our model can also be used to study planets with more complex structures, provided that we use modified Love numbers.
In addition, we can generalize it to other rheologies.
In order to determine the specific deformation law that replaces expression (\ref{131021e}), one just needs to know how the strain is related to the stress in each case (see appendix~\ref{apenA}).
However, in the case of complex rheologies, 
this method may lead  to a differential system of very high order, that is less easy to handle. 

The inclusion of general relativity or the effect from an oblate star would result in a correction to the precession rate of the pericenter, $\dot \varpi$ \citep[e.g.,][]{Correia_etal_2011}. 
Since the rotation rate is defined as $\om = d \ttheta / dt = \dot \theta - \dot \varpi$ (Eq.\,\ref{131105b}), these effects do not modify the conclusions and the results reported in this paper.

\begin{acknowledgements}
We acknowledge support from the PNP-CNRS, CS of Paris Observatory, PICS05998 France-Portugal program, FCT-Portugal (PEst-C/CTM/LA0025/2011), and FAPESP-Brazil (2009/16900-5, 2012/13731-0).
\end{acknowledgements}

\bibliographystyle{aa}
\bibliography{correia}

\begin{thebibliography}{64}
\expandafter\ifx\csname natexlab\endcsname\relax\def\natexlab#1{#1}\fi

\bibitem[{{Adel Sharaf} \& {Hassan Selim}(2010)}]{Sharaf_Selim_2010}
{Adel Sharaf}, M. \& {Hassan Selim}, H. 2010, Research in Astronomy and
  Astrophysics, 10, 1298

\bibitem[{{Alexander}(1973)}]{Alexander_1973}
{Alexander}, M.~E. 1973, \apss, 23, 459

\bibitem[{{Alibert} {et~al.}(2005){Alibert}, {Mordasini}, {Benz}, \&
  {Winisdoerffer}}]{Alibert_etal_2005}
{Alibert}, Y., {Mordasini}, C., {Benz}, W., \& {Winisdoerffer}, C. 2005, \aap,
  434, 343

\bibitem[{{Andrade}(1910)}]{Andrade_1910}
{Andrade}, E.~N.~d.~C. 1910, Proc. R. Soc. Lond. A, 84, 1

\bibitem[{{Auclair-Desrotour} {et~al.}(2014){Auclair-Desrotour}, {Le
  Poncin-Lafitte}, \& {Mathis}}]{Auclair-Desrotour_etal_2014}
{Auclair-Desrotour}, P., {Le Poncin-Lafitte}, C., \& {Mathis}, S. 2014, \aap,
  561, L7

\bibitem[{{Bakos} {et~al.}(2012){Bakos}, {Hartman}, {Torres}, {B{\'e}ky},
  {Latham}, {Buchhave}, {Csubry}, {Kov{\'a}cs}, {Bieryla}, {Quinn},
  {Szklen{\'a}r}, {Esquerdo}, {Shporer}, {Noyes}, {Fischer}, {Johnson},
  {Howard}, {Marcy}, {Sato}, {Penev}, {Everett}, {Sasselov}, {F{\H u}r{\'e}sz},
  {Stefanik}, {L{\'a}z{\'a}r}, {Papp}, \& {S{\'a}ri}}]{Bakos_etal_2012}
{Bakos}, G.~{\'A}., {Hartman}, J.~D., {Torres}, G., {et~al.} 2012, \aj, 144, 19

\bibitem[{{Beaug{\'e}} \& {Nesvorn{\'y}}(2012)}]{Beauge_Nesvorny_2012}
{Beaug{\'e}}, C. \& {Nesvorn{\'y}}, D. 2012, \apj, 751, 119

\bibitem[{{Colombo}(1965)}]{Colombo_1965}
{Colombo}, G. 1965, \nat, 208, 575

\bibitem[{{Correia}(2009)}]{Correia_2009}
{Correia}, A.~C.~M. 2009, \apjl, 704, L1

\bibitem[{{Correia} {et~al.}(2011){Correia}, {Laskar}, {Farago}, \&
  {Bou{\'e}}}]{Correia_etal_2011}
{Correia}, A.~C.~M., {Laskar}, J., {Farago}, F., \& {Bou{\'e}}, G. 2011,
  Celestial Mechanics and Dynamical Astronomy, 111, 105

\bibitem[{{Correia} \& {Rodr{\'{\i}}guez}(2013)}]{Correia_Rodriguez_2013}
{Correia}, A.~C.~M. \& {Rodr{\'{\i}}guez}, A. 2013, \apj, 767, 128

\bibitem[{{Darwin}(1879)}]{Darwin_1879a}
{Darwin}, G.~H. 1879, Philos. Trans. R. Soc. London, 170, 1

\bibitem[{{Darwin}(1880)}]{Darwin_1880}
{Darwin}, G.~H. 1880, Philos. Trans. R. Soc. London, 171, 713

\bibitem[{{Dawson} \& {Murray-Clay}(2013)}]{Dawson_Murray-Clay_2013}
{Dawson}, R.~I. \& {Murray-Clay}, R.~A. 2013, \apjl, 767, L24

\bibitem[{{Dickey} {et~al.}(1994){Dickey}, {Bender}, {Faller}, {Newhall},
  {Ricklefs}, {Ries}, {Shelus}, {Veillet}, {Whipple}, {Wiant}, {Williams}, \&
  {Yoder}}]{Dickey_etal_1994}
{Dickey}, J.~O., {Bender}, P.~L., {Faller}, J.~E., {et~al.} 1994, Science, 265,
  482

\bibitem[{{Efroimsky}(2012)}]{Efroimsky_2012}
{Efroimsky}, M. 2012, Celestial Mechanics and Dynamical Astronomy, 112, 283

\bibitem[{{Efroimsky} \& {Williams}(2009)}]{Efroimsky_Williams_2009}
{Efroimsky}, M. \& {Williams}, J.~G. 2009, Celestial Mechanics and Dynamical
  Astronomy, 104, 257

\bibitem[{{Favier} {et~al.}(2014){Favier}, {Barker}, {Baruteau}, \&
  {Ogilvie}}]{Favier_etal_2014}
{Favier}, B., {Barker}, A.~J., {Baruteau}, C., \& {Ogilvie}, G.~I. 2014,
  \mnras, 439, 845

\bibitem[{{Ferraz-Mello}(2013)}]{Ferraz-Mello_2013}
{Ferraz-Mello}, S. 2013, Celestial Mechanics and Dynamical Astronomy, 116, 109

\bibitem[{{Gillon} {et~al.}(2012){Gillon}, {Demory}, {Benneke}, {Valencia},
  {Deming}, {Seager}, {Lovis}, {Mayor}, {Pepe}, {Queloz}, {S{\'e}gransan}, \&
  {Udry}}]{Gillon_etal_2012b}
{Gillon}, M., {Demory}, B.-O., {Benneke}, B., {et~al.} 2012, \aap, 539, A28

\bibitem[{{Goldreich} \& {Peale}(1966)}]{Goldreich_Peale_1966}
{Goldreich}, P. \& {Peale}, S. 1966, \aj, 71, 425

\bibitem[{{Guenel} {et~al.}(2014){Guenel}, {Mathis}, \&
  {Remus}}]{Guenel_etal_2014}
{Guenel}, M., {Mathis}, S., \& {Remus}, F. 2014, \aap, 566, L9

\bibitem[{{Guillot}(1999)}]{Guillot_1999}
{Guillot}, T. 1999, Science, 286, 72

\bibitem[{{H{\'e}brard} {et~al.}(2010){H{\'e}brard}, {D{\'e}sert},
  {D{\'{\i}}az}, {Boisse}, {Bouchy}, {Lecavelier Des Etangs}, {Moutou},
  {Ehrenreich}, {Arnold}, {Bonfils}, {Delfosse}, {Desort}, {Eggenberger},
  {Forveille}, {Gregorio}, {Lagrange}, {Lovis}, {Pepe}, {Perrier}, {Pont},
  {Queloz}, {Santerne}, {Santos}, {S{\'e}gransan}, {Sing}, {Udry}, \&
  {Vidal-Madjar}}]{Hebrard_etal_2010b}
{H{\'e}brard}, G., {D{\'e}sert}, J.-M., {D{\'{\i}}az}, R.~F., {et~al.} 2010,
  \aap, 516, A95

\bibitem[{{Henning} {et~al.}(2009){Henning}, {O'Connell}, \&
  {Sasselov}}]{Henning_etal_2009}
{Henning}, W.~G., {O'Connell}, R.~J., \& {Sasselov}, D.~D. 2009, \apj, 707,
  1000

\bibitem[{{Howard} {et~al.}(2013){Howard}, {Sanchis-Ojeda}, {Marcy}, {Johnson},
  {Winn}, {Isaacson}, {Fischer}, {Fulton}, {Sinukoff}, \&
  {Fortney}}]{Howard_etal_2013}
{Howard}, A.~W., {Sanchis-Ojeda}, R., {Marcy}, G.~W., {et~al.} 2013, ArXiv
  e-prints

\bibitem[{{Hughes}(1981)}]{Hughes_1981}
{Hughes}, S. 1981, Celestial Mechanics, 25, 101

\bibitem[{{Hut}(1980)}]{Hut_1980}
{Hut}, P. 1980, \aap, 92, 167

\bibitem[{{Ida} \& {Lin}(2008)}]{Ida_Lin_2008}
{Ida}, S. \& {Lin}, D.~N.~C. 2008, \apj, 685, 584

\bibitem[{{Jeffreys}(1976)}]{Jeffreys_1976}
{Jeffreys}, H. 1976, {The earth. Its origin, history and physical
  constitution.}

\bibitem[{{Kant}(1754)}]{Kant_1754}
{Kant}, I. 1754, {Kant's cosmogony: As in his essay on the retardation of the
  rotation of the earth and his Natural history and theory of the heavens}

\bibitem[{{Kaula}(1964)}]{Kaula_1964}
{Kaula}, W.~M. 1964, \rg, 2, 661

\bibitem[{{Kelvin}(1863{\natexlab{a}})}]{Kelvin_1863b}
{Kelvin}, W.~T.~B. 1863{\natexlab{a}}, Philosophical Transactions of the Royal
  Society of London, 153, 583

\bibitem[{{Kelvin}(1863{\natexlab{b}})}]{Kelvin_1863a}
{Kelvin}, W.~T.~B. 1863{\natexlab{b}}, Philosophical Transactions of the Royal
  Society of London, 153, 573

\bibitem[{{Konopliv} {et~al.}(2006){Konopliv}, {Yoder}, {Standish}, {Yuan}, \&
  {Sjogren}}]{Konopliv_etal_2006}
{Konopliv}, A.~S., {Yoder}, C.~F., {Standish}, E.~M., {Yuan}, D.-N., \&
  {Sjogren}, W.~L. 2006, \icarus, 182, 23

\bibitem[{{Lainey} {et~al.}(2009){Lainey}, {Arlot}, {Karatekin}, \& {van
  Hoolst}}]{Lainey_etal_2009}
{Lainey}, V., {Arlot}, J.-E., {Karatekin}, {\"O}., \& {van Hoolst}, T. 2009,
  Nature, 459, 957

\bibitem[{{Lainey} {et~al.}(2007){Lainey}, {Dehant}, \&
  {P{\"a}tzold}}]{Lainey_etal_2007}
{Lainey}, V., {Dehant}, V., \& {P{\"a}tzold}, M. 2007, \aap, 465, 1075

\bibitem[{{Lainey} {et~al.}(2012){Lainey}, {Karatekin}, {Desmars}, {Charnoz},
  {Arlot}, {Emelyanov}, {Le Poncin-Lafitte}, {Mathis}, {Remus}, {Tobie}, \&
  {Zahn}}]{Lainey_etal_2012}
{Lainey}, V., {Karatekin}, {\"O}., {Desmars}, J., {et~al.} 2012, \apj, 752, 14

\bibitem[{{Laskar} \& {Bou{\'e}}(2010)}]{Laskar_Boue_2010}
{Laskar}, J. \& {Bou{\'e}}, G. 2010, \aap, 522, A60

\bibitem[{{Laskar} {et~al.}(2012){Laskar}, {Bou{\'e}}, \&
  {Correia}}]{Laskar_etal_2012}
{Laskar}, J., {Bou{\'e}}, G., \& {Correia}, A.~C.~M. 2012, \aap, 538, A105

\bibitem[{{Lin} {et~al.}(1996){Lin}, {Bodenheimer}, \&
  {Richardson}}]{Lin_etal_1996}
{Lin}, D.~N.~C., {Bodenheimer}, P., \& {Richardson}, D.~C. 1996, \nat, 380, 606

\bibitem[{{Love}(1911)}]{Love_1911}
{Love}, A.~E.~H. 1911, {Some Problems of Geodynamics}

\bibitem[{{MacDonald}(1964)}]{MacDonald_1964}
{MacDonald}, G.~J.~F. 1964, Revs. Geophys., 2, 467

\bibitem[{{Makarov} \& {Efroimsky}(2013)}]{Makarov_Efroimsky_2013}
{Makarov}, V.~V. \& {Efroimsky}, M. 2013, \apj, 764, 27

\bibitem[{{Mathews} {et~al.}(1995){Mathews}, {Buffet}, \&
  {Shapiro}}]{Mathews_etal_1995}
{Mathews}, P.~M., {Buffet}, B.~A., \& {Shapiro}, I.~I. 1995, \grl, 22, 579

\bibitem[{{Mathis} \& {Le Poncin-Lafitte}(2009)}]{Mathis_Poncin-Lafitte_2009}
{Mathis}, S. \& {Le Poncin-Lafitte}, C. 2009, \aap, 497, 889

\bibitem[{{Mignard}(1979)}]{Mignard_1979}
{Mignard}, F. 1979, Moon and Planets, 20, 301

\bibitem[{{Mordasini} {et~al.}(2009){Mordasini}, {Alibert}, \&
  {Benz}}]{Mordasini_etal_2009a}
{Mordasini}, C., {Alibert}, Y., \& {Benz}, W. 2009, \aap, 501, 1139

\bibitem[{{Newton}(1760)}]{Newton_1760}
{Newton}, I. 1760, {Philosophi\ae Naturalis Principia Mathematica}, ed.
  {Newton, I.}

\bibitem[{{Ogilvie} \& {Lesur}(2012)}]{Ogilvie_Lesur_2012}
{Ogilvie}, G.~I. \& {Lesur}, G. 2012, \mnras, 422, 1975

\bibitem[{{Ogilvie} \& {Lin}(2004)}]{Ogilvie_Lin_2004}
{Ogilvie}, G.~I. \& {Lin}, D.~N.~C. 2004, \apj, 610, 477

\bibitem[{{Rasio} \& {Ford}(1996)}]{Rasio_Ford_1996}
{Rasio}, F.~A. \& {Ford}, E.~B. 1996, Science, 274, 954

\bibitem[{{Ray} {et~al.}(2001){Ray}, {Eanes}, \& {Lemoine}}]{Ray_etal_2001}
{Ray}, R.~D., {Eanes}, R.~J., \& {Lemoine}, F.~G. 2001, Geophysical Journal
  International, 144, 471

\bibitem[{{Remus} {et~al.}(2012{\natexlab{a}}){Remus}, {Mathis}, \&
  {Zahn}}]{Remus_etal_2012b}
{Remus}, F., {Mathis}, S., \& {Zahn}, J.-P. 2012{\natexlab{a}}, \aap, 544, A132

\bibitem[{{Remus} {et~al.}(2012{\natexlab{b}}){Remus}, {Mathis}, {Zahn}, \&
  {Lainey}}]{Remus_etal_2012a}
{Remus}, F., {Mathis}, S., {Zahn}, J.-P., \& {Lainey}, V. 2012{\natexlab{b}},
  \aap, 541, A165

\bibitem[{{Rodr{\'{\i}}guez} {et~al.}(2012){Rodr{\'{\i}}guez}, {Callegari},
  {Michtchenko}, \& {Hussmann}}]{Rodriguez_etal_2012}
{Rodr{\'{\i}}guez}, A., {Callegari}, N., {Michtchenko}, T.~A., \& {Hussmann},
  H. 2012, \mnras, 427, 2239

\bibitem[{{Singer}(1968)}]{Singer_1968}
{Singer}, S.~F. 1968, \gjras, 15, 205

\bibitem[{{Socrates} {et~al.}(2012){Socrates}, {Katz}, \&
  {Dong}}]{Socrates_etal_2013}
{Socrates}, A., {Katz}, B., \& {Dong}, S. 2012, ArXiv e-prints

\bibitem[{{Storch} \& {Lai}(2014)}]{Storch_Lai_2014}
{Storch}, N.~I. \& {Lai}, D. 2014, \mnras, 438, 1526

\bibitem[{{Turcotte} \& {Schubert}(2002)}]{Turcotte_Schubert_2002}
{Turcotte}, D.~L. \& {Schubert}, G. 2002, {Geodynamics}

\bibitem[{{Valencia} {et~al.}(2007){Valencia}, {Sasselov}, \&
  {O'Connell}}]{Valencia_etal_2007a}
{Valencia}, D., {Sasselov}, D.~D., \& {O'Connell}, R.~J. 2007, \apj, 656, 545

\bibitem[{{Wu} \& {Murray}(2003)}]{Wu_Murray_2003}
{Wu}, Y. \& {Murray}, N. 2003, \apj, 589, 605

\bibitem[{{Yoder}(1995)}]{Yoder_1995cnt}
{Yoder}, C.~F. 1995, in Global Earth Physics: A Handbook of Physical Constants
  (American Geophysical Union, Washington D.C), 1--31

\bibitem[{{Zahn}(1977)}]{Zahn_1977}
{Zahn}, J.-P. 1977, \aap, 57, 383

\end{thebibliography}

\appendix

\section{Fourier domain}
\label{apenA}
A planet with mass $\m$ and radius $R$ placed in a static external quadrupolar gravitational potential $V_p(\vr)$ deforms itself until it reaches a hydrostatic equilibrium. In turn, this deformation generates an additional gravitational potential $V'(\vr)$. For a homogeneous incompressible elastic solid body, the additional potential is given on the planet's surface by \citep{Kelvin_1863b, Kelvin_1863a, Love_1911}
\be
V'_\eq(\vR) = k_{2} V_p(\vR)\ ,
\label{eq.love1}
\ee
where $k_{2}$ is the second Love number associated with a static perturbing potential. It is related to the rigidity $\mue$ of the planet through \citep{Kelvin_1863b, Kelvin_1863a, Love_1911}
\be
k_{2} = \frac{3}{2}\left(1+\frac{19\mue}{2g\rho R}\right)^{-1}\ .
\ee
According to the {\em correspondence principle}, in the more general case where the external gravitational potential is not static, expression (\ref{eq.love1}) remains the same in the Fourier domain \citep[and references therein]{Darwin_1879a, Efroimsky_2012}, and thus
\be
\hat V'(\vR, \sig) = \hat k_2(\sig) \hat V_p(\vR, \sig)\ ,
\label{eq.TFlove}
\ee
where $\sig$ is a frequency, and
\be
\hat k_2(\sig) = \frac{3}{2}\left(1+
\frac{19\hat\mue(\sig)}{2g\rho R}\right)^{-1}\ .
\label{eq.TFk2}
\ee
In the case of a viscoelastic material characterized by Maxwell rheology -- represented at the mesoscopic scale by a spring with elastic modulus $\mue$ and a damper with viscosity $\nue$ put in parallel -- the strain $\varepsilon$ is related to the stress $\sigma$ through the differential equation
\be
\frac{d\varepsilon}{dt} = \frac{1}{\mue}\frac{d\sigma}{dt} +
\frac{\sigma}{\nue}\ .
\label{eq.rheology}
\ee
The expression of $\hat \mue(\sig) \equiv \hat \sigma(\sig)/\hat \varepsilon(\sig)$ is then obtained by performing a Fourier transform on (\ref{eq.rheology}), which leads to
\be
\hat \mue(\sig)  = \mue \frac{\ii \taua \sig}{1+\ii \taua \sig}\ ,
\label{eq.TFmu}
\ee
with $\taua=\nue/\mue$ being the Maxwell relaxation time.  Equations (\ref{eq.TFlove}), (\ref{eq.TFk2}), and (\ref{eq.TFmu}), are then combined together to get \citep[e.g.,][]{Henning_etal_2009}
\be
\hat V'(\vR, \sig) = \kf \frac{1+\ii \taua \sig}{1+\ii \taub \sig} \hat V_p(\vR, \sig)\ ,
\label{eq.transf}
\ee
where $\kf=3/2$ is the fluid second Love number and
\be
\taub = \taua + \frac{19\nue}{2g\rho R}\ .
\ee
In a last step, we convert (\ref{eq.transf}) into a differential equation in the temporal domain by multiplying both sides by $1+\ii\taub\sig$ and substituting ``$\ii\sig$'' by ``$d/dt$''. The result is
\be
V' + \taub \dot V' = \kf \left(V_p+\taua\dot V_p\right)\ .
\ee

\section{Linear model}
\label{apenB}
The secular evolution of the rotation rate is given by (Eq.\,\ref{ddotthetaM})
\be
\left\langle \ddot \theta \right\rangle_M = - \kf \frac{3 G \M^2 R^5}{2 C a^6} \sum^{+\infty}_{k=-\infty}  \left[ X_k^{-3,2} (e) \right]^2 \frac{(\taub-\taua)\sig_k}{1 + \taub^2\sig_k^2} \ ,
\label{eq.torque}
\ee
In the low frequency regime ($\taub \sig_k\ll 1$), using  $\sig_k=2\om-kn$ and $\Delta \tau = \taub-\taua$, we can rewrite previous expression as
\begin{eqnarray}
\left\langle \ddot \theta \right\rangle_M &=& - \kf \frac{3 G \M^2 R^5}{2 C a^6} \sum^{+\infty}_{k=-\infty}  \left[ X_k^{-3,2} (e) \right]^2 \Delta \tau \, (2\om-kn)  \crm
&=& - \kf \frac{3 G \M^2 R^5}{C a^6} n \Delta \tau
\left(f_1(e) \frac{\om}{n} - f_2(e)\right)\ ,
\label{eq.torque2}
\end{eqnarray}
where
\be
f_1(e) = \sum^{+\infty}_{k=-\infty} \left[ X_k^{-3,2} (e) \right]^2 \ ,
\ee
and
\be
f_2(e) = \frac{1}{2}  \sum^{+\infty}_{k=-\infty} k \left[ X_k^{-3,2} (e) \right]^2\ .
\ee
We let 
\be
F = \left(\frac{a}{r}\right)^3 \ep^{\ii 2 \av}
\quad \mathrm{and} \quad
\conj{F} = \left(\frac{a}{r}\right)^3 \ep^{-\ii 2 \av} \ . \llabel{140122a}
\ee
Using the definition of the Hansen coefficients (Eq.\,\ref{061120gb}), we get
\begin{eqnarray}
f_1(e) \!&=&\! \left\langle F \conj{F} \right\rangle_M
    = \left\langle \left(\frac{a}{r}\right)^6 \right\rangle_M \crm
    \!&=&\! X_0^{-6,0}(e)
    = \frac{1+3e^2+\frac{3}{8}e^4}{(1-e^2)^{9/2}}
\label{eq.S0}
\end{eqnarray}
and
\begin{eqnarray}
f_2(e) \!&=&\! \frac{1}{2}   \left\langle \frac{dF}{dM} \conj{F} \right\rangle_M
    = \left\langle  \left(\frac{a}{r}\right)^8 \sqrt{1-e^2} \right\rangle_M   \crm
    \!&=&\! X_0^{-8,0}(e)  \sqrt{1-e^2} 
    = \frac{1+\frac{15}{2}e^2+\frac{45}{8}e^4+\frac{5}{16}e^6}{(1-e^2)^6}\ .
\label{eq.S1}
\end{eqnarray}
The expressions for  $X_0^{-6,0}(e)$ and $X_0^{-8,0}(e) $ can be found in
\citep[e.g.,][]{Laskar_Boue_2010}.

\section{Hansen coefficients}
\label{appendix.Hansen}

The construction of Figures \ref{Fig1c} and \ref{Fig2}, requires the computation of many Hansen coefficients
(Eqs.\ref{torque} and \ref{131105m}). These coefficients can be efficiently computed using Fourier techniques.
The functions $F^{l,m} = \left(\frac{r}{a}\right)^l\ep^{\ii m \av}$ are
$2\pi$-periodic functions of mean anomaly $M$. They can thus be
developed in Fourier series with integer frequencies $k\in\mathbb{Z}$.
The corresponding Fourier coefficients are precisely the Hansen
coefficients $X_k^{l,m} (e)$. Using this property, it is possible to get
at once a numerical estimate of $2^N+1$ Hansen coefficients for 
$-K\leq k \leq K$ with $K=2^{N-1}$ by means of a Fast Fourier Transform (FFT) 
\citep{Sharaf_Selim_2010}. For that purpose, each function $F^{l,m}(M)$
has to be evenly sampled in an array {\ttfamily y[0:$2^N-1$]} of 
size $2^N-1$ with {\ttfamily y[j] := $F^{l,m}(M_j)$} where
\be
M_j = j \frac{2\pi}{2^N}
\ee
and $j=0,\ldots,2^N-1$. Then, the output {\ttfamily y\_FFT[0:$2^N-1$]} of
the FFT algorithm contains $2^N-1$ Hansen coefficients. If the $j$th
term {\ttfamily y\_FFT[j]} is associated with the frequency $\omega_j = j$, the Hansen
coefficients are sorted in the following order
\be
k = 0, 1, \ldots, K, 1-K, \ldots, -2, -1\ .
\ee
The last coefficient is deduced from $X^{l,m}_{-K}(e) = X^{l,m}_{K}(e)$.
In Table~\ref{TabX} we list the Hansen coefficients $X_k^{-3,2} (e) $ up to $ e^6 $, that appear often throughout this paper.

\begin{table}[h!]
\caption{Hansen coefficients $X_k^{-3,2} (e) $ up to $ e^6 $.
The exact expression of these coefficients is given by $  X_k^{-3,2} (e)  =
\frac{1}{\pi} \int_0^\pi \left( \frac{r}{a} \right)^{-3} \exp(\ii \, 2 \lv)
\exp(\ii \, k M) \, d M $. \llabel{TabX} } 
\begin{center}
{ \begin{tabular}{| c |ccccccc|} \hline 
$  k  $ & \multicolumn{7}{c|}{$ \phantom{\Frac{1}{1}} X_k^{-3,2} (e) \phantom{\Frac{1}{1}} $} \\ \hline 
$-$4 & & & & & & & $\Frac{4}{45} e^6$ \\ 
$-$3 & & & & & & $\Frac{81}{1280} e^5$ & \\ %
$-$2 & & & & & $\Frac{1}{24} e^4$ & $+$ & $\Frac{7}{240} e^6$ \\ 
$-$1 & & & & $\Frac{1}{48} e^3$ & $+$ & $\Frac{11}{768} e^5$ & \\ %
1 & $-$ & $\Frac{1}{2} e$ & $+$ & $\Frac{1}{16} e^3$ & $-$ & $\Frac{5}{384} e^5$ & \\ 
2 & 1 & $-$ & $\Frac{5}{2} e^2$ & $+$ & $\Frac{13}{16} e^4$ & $-$ & $\Frac{35}{288} e^6$ \\ 
3 & & $\Frac{7}{2} e$ & $-$ & $\Frac{123}{16} e^3$ & $+$ & $\Frac{489}{128} e^5$ & \\ 
4 & & & $\Frac{17}{2} e^2$ & $-$ & $\Frac{115}{6} e^4$ & $+$ & $\Frac{601}{48} e^6$ \\ 
5 & & & & $\Frac{845}{48} e^3$ & $-$ & $\Frac{32525}{768} e^5$ & \\ 
6 & & & & & $\Frac{533}{16} e^4$ & $-$ & $\Frac{13827}{160} e^6$ \\ 
7 & & & & & & $\Frac{228347}{3840} e^5$ & \\ 
8 & & & & & & & $\Frac{73369}{720} e^6$ \\ \hline 
\end{tabular} 
}
\end{center}
\end{table}

\end{document}